\documentclass[12pt,a4paper]{article}
\usepackage{jheppub}  
\usepackage{amssymb} 
\usepackage{amsmath}
\usepackage{mathtools}
\usepackage{amsfonts}    
\usepackage{dsfont}
\usepackage{pdfpages}
\usepackage{verbatim}
\usepackage{tensor}
\usepackage{mathrsfs}
\usepackage{textgreek} 
\usepackage[mathscr]{euscript}
\usepackage{tikz}
\usetikzlibrary{patterns,decorations.pathreplacing}

\newcommand{\ba}{\begin{align}}

\newcommand{\be}{\begin{equation}}
\newcommand{\ee}{\end{equation}}
\def\bd{\begin{tikzpicture}}
\def\ed{\end{tikzpicture}}

\renewcommand\Re{\mathop{\text{Re}}}
\newcommand\Res{\mathop{\text{Res}}}

\allowdisplaybreaks[1]

\newcommand{\G}{\mathrm{G}}
\newcommand{\lgamma}{\text{\textgamma}}

\title{A perturbative CFT dual for pure NS-NS AdS$_{\mathbf{3}}$ strings}

\author{Lorenz Eberhardt} 
\affiliation{School of Natural Sciences, Institute for Advanced Study, \\
\hspace*{0.3cm}Einstein Drive 1, Princeton,  NJ 08540, USA}
\emailAdd{elorenz@ias.edu}

\abstract{We construct a CFT dual to string theory on $\mathrm{AdS}_3$ with pure NS-NS flux. It is given by a symmetric orbifold of a linear dilaton theory deformed by a marginal operator from the twist-2 sector. We compute two- and three-point functions on the CFT side to 4th order in conformal perturbation theory at large $N$. They agree with the string computation at genus 0, thus providing ample evidence for a duality. We also show that the full spectra of both short and long strings on the CFT and the string side match. The duality should be understood as perturbative in $N^{-1}$.}

\begin{document}

\maketitle

%make math in all titles bold
\makeatletter
\g@addto@macro\bfseries{\boldmath}
\makeatother
%end code

%%%%%%%%%%%%%%%%%%%%%%%%%%%%%%%%%%%%%%%%%%%%%%%%%%%%%%%%%%%%%%%

\section{Introduction}
String theory on $\mathrm{AdS}_3$ with pure NS-NS flux is among the string backgrounds with the best computational control. The theory is described by the $\mathrm{SL}(2,\mathbb{R})_k$ WZW-model on the worldsheet. Its spectrum has been understood for a long time \cite{Balog:1988jb, Petropoulos:1989fc, Hwang:1990aq, Henningson:1991jc, Gawedzki:1991yu, Bars:1995mf,  Evans:1998qu, Giveon:1998ns, deBoer:1998gyt, Kutasov:1999xu,  Maldacena:2000hw, Maldacena:2000kv}, while correlation functions on the sphere were discussed in \cite{Teschner:1997ft, Teschner:1999ug, Giribet:1999ft, Giribet:2000fy, Maldacena:2001km,  Giribet:2001ft,  Ribault:2005wp, Giribet:2005ix, Ribault:2005ms, Hikida:2007tq,  Giribet:2005mc, Minces:2005nb,Iguri:2007af,Baron:2008qf,Iguri:2009cf,Giribet:2011xf,Cagnacci:2013ufa,Giribet:2015oiy,Giribet:2019new,Hikida:2020kil, Dei:2021xgh, Dei:2021yom}. Compared to compact WZW-models, the theory is significantly more complicated, mainly because of the existence of spectrally flowed representations of the affine algebra. Spectrally flowed vertex operators correspond to winding strings in spacetime and contain most of the interesting physics in the background.

Given the good technical control on the worldsheet side, it is somewhat ironic that pure NS-NS backgrounds are among the least understood in the context of the AdS/CFT duality \cite{Maldacena:1997re}. For superstrings on $\mathrm{AdS}_3 \times \mathrm{S}^3 \times \mathbb{T}^4$, it is conjectured that the dual CFT is a singular point in the moduli space of the symmetric orbifold of $\mathbb{T}^4$ \cite{Seiberg:1999xz}. This is however not very useful in practice, since one is restricted to the comparison of BPS protected quantities, which has successfully been done in \cite{Strominger:1996sh, deBoer:1998us, Maldacena:1999bp, Gaberdiel:2007vu, Dabholkar:2007ey, Pakman:2007hn, Giribet:2007wp, Cardona:2009hk}. 

One of the main reasons why the CFT duals of pure NS-NS $\mathrm{AdS}_3$ backgrounds have remained elusive is the presence of so-called long strings in the spectrum. The Hilbert space has a continuous sector of delta-function normalizable strings that can expand towards the boundary of $\mathrm{AdS}_3$. Hence the dual CFT cannot be a compact CFT.

Progress was made in the last couple of years for a particular limit of string theory, where the size of $\mathrm{AdS}_3$ becomes of the order of the string and hence the string is tensionless. This corresponds to $k=1$ for the level of the (supersymmetric) WZW model describing $\mathrm{AdS}_3$. In this tensionless regime, the continuous sector of the Hilbert space disappears and
one finds that the dual CFT is simply the symmetric orbifold of $\mathbb{T}^4$ \cite{Gaberdiel:2018rqv, Eberhardt:2018ouy, Eberhardt:2019ywk, Eberhardt:2020akk, Dei:2020zui, Knighton:2020kuh, Gaberdiel:2020ycd, Eberhardt:2020bgq, Eberhardt:2021jvj}. There is a second class of backgrounds with recent progress \cite{Balthazar:2021xeh, Martinec:2021vpk}. These backgrounds have $k<1$ and describe the transverse space to NS5-branes and fundamental strings. Due to their stringy nature, the number of target space dimensions is less than 10. In this case, the dual CFT can be described conjecturally as a symmetric orbifold theory deformed by a certain `wall operator'.
\bigskip

These developments motivate us to reconsider the generic pure NS-NS background. The worldsheet theory makes a prediction for the correlation functions of vertex operators. As such, one may wonder whether we can just use it as the \emph{definition} of the dual spacetime CFT, at least perturbatively in the string coupling. If we are able to construct a CFT that has the correct three-point functions, then we have effectively \emph{derived} the spacetime CFT from the worldsheet.

In this paper, we show that this possibility is indeed realized. For technical simplicity, we mainly focus on the bosonic string. Even though it suffers from the usual instability due to the tachyon, one can still study the AdS/CFT duality at tree-level, where it is perfectly well-defined.
We construct a deformed symmetric orbifold CFT that reproduces all worldsheet two- and three-point functions at least up to 4th order in the deformation parameter. The symmetric orbifold in question is
\be 
\mathrm{Sym}^N(\mathbb{R}_Q \times X)\ , \label{eq:intro symmetric orbifold}
\ee 
where $\mathbb{R}_Q$ is a linear dilaton direction and $X$ is the internal CFT of the compactification. The CFT is then deformed by a twist-2 field dressed by a free boson vertex operator in the linear dilaton direction. Explicitly, the (non-normalizable) marginal operator is $\mathrm{e}^{-\sqrt{k-2} \phi} \sigma_2$ with $\phi$ the linear dilaton and $\sigma_2$ the twist-2 field. In some ways, this deformation is similar to how Liouville theory can be thought of as a deformed linear dilaton theory, but it gets complicated by the non-trivial interaction with the twisted sectors.
We will use similar technology as in the construction of Liouville theory from the Coulomb gas formalism \cite{Dotsenko:1984nm, Dotsenko:1984ad, Dorn:1994xn, Zamolodchikov:1995aa} to compute correlation functions of this theory. We use this modified Coulomb gas formalism to reproduce the known string theory three-point functions up to 4th order in the deformation parameter $\mu$. The matching is very non-trivial and relies on many surprising identities for branched coverings that are needed to compute the correlators in the symmetric product orbifold. 
We simply check these identities numerically in the ancillary \texttt{Mathematica} notebook for a large number of cases. However, the general matching of these correlators seems a mathematically well-defined problem and it should in principle be possible to give a proof that the correlators agree to all orders in conformal perturbation series.

We also check other predictions of the duality, some of which go beyond perturbation theory in the deforming parameter $\mu$. We check that the pole structure of the three-point functions agrees. There are two types of poles -- LSZ poles and bulk poles, which we discuss in detail \cite{Aharony:2004xn}. We also show that the three-point function with an inserted marginal operator reduces to the the $\mu$-derivative of the two-point function on the worldsheet. This is a prediction from the conjectured dual CFT, that is highly non-trivial from the worldsheet perspective. 

We match the three-point functions first for the continuous sector of the theory. The discrete spectrum can then be read off from the poles in the correlation functions. We argue that some of these poles lead to normalizable states, which allows us to conclude that the full spectrum of delta-function normalizable and normalizable operators of the two theories agrees.

The constructed dual CFT has some peculiar properties. The main difficulty is that its construction is somewhat ad hoc, since we can only access it in perturbation theory that is not particularly well-behaved because the deforming operator is non-normalizable. For the perturbation theory to make sense to all orders, one has to consider the limit $N \to \infty$ in the symmetric product orbifold. Thus, we currently only understand the proposed CFT as a large $N$ CFT. We are in particular currently unable to say anything in this CFT that goes beyond string perturbation theory. 
Another interesting consequence of the large $N$ limit is that $N$ only appears as a normalization of the partition function and the vertex operators in this limit and hence can be completely eliminated from the theory. The CFT ultimately depends only on the deformation parameter $\mu$ that in turn gets related to the string coupling. The proposed dual description hence does not depend on $N$ anymore (as long as $N$ is asymptotically large) and so does not have any particular central charge. It should rather be thought of as a `grand canonical ensemble' of CFTs, a point of view that was advocated in \cite{Kim:2015gak, Eberhardt:2020bgq, Eberhardt:2021jvj}.
\medskip

Our proposal improves upon previous work \cite{Eberhardt:2019qcl, Dei:2019osr}, see also \cite{Seiberg:1999xz, Argurio:2000tb}, where the symmetric orbifold \eqref{eq:intro symmetric orbifold} was proposed as describing the long string sector of $\mathrm{AdS}_3$. It was also proposed that the theory should be deformed by a marginal deformation to something akin to the symmetric orbifold of Liouville theory. The proposal missed however that the marginal operator should include the twisted sectors in a non-trivial way. In other words, the marginal operator in these works was assumed to be of the form $\mathrm{e}^{\alpha \phi}$, where $\phi$ is the linear dilaton direction and $\alpha$ is chosen such that the operator is marginal, whereas we show here that it should be $\mathrm{e}^{\alpha \phi} \sigma_2$, where $\sigma_2$ is the twist-2 field. All other essential ingredients were present in these works, in particular the spectrum of long strings is independent of the precise marginal operator and is correctly reproduced in both CFTs.
\bigskip

This paper is organized as follows. In Section~\ref{sec:setup}, we introduce and review the two relevant theories of interest: string worldsheet theory on $\mathrm{AdS}_3$ and the deformed symmetric product orbifold. Most of this material is known, but is scattered across the literature. Section~\ref{sec:matching} contains our main technical computations that give strong evidence that the two- and three-point functions in both theories agree. We then discuss poles in the correlation functions in Section~\ref{sec:poles and short strings} and extract the spectrum of discrete normalizable operators in the dual CFT. In Section~\ref{sec:grand canonical ensemble}, we explain the subtleties of our proposal that require us to consider the large $N$ limit. Finally, we generalize our proposal to superstrings in Section~\ref{sec:superstrings}. While we do not perform detailed calculations in this case, the structure is sufficiently similar that we can be confident in the duality. We end in Section~\ref{sec:discussion} with some open problems and future directions.

\medskip
While this paper was in preparation \cite{Balthazar:2021xeh, Martinec:2021vpk} appeared, which contain overlapping ideas.

\section{Setup} \label{sec:setup}
In this section, we explain the two theories whose duality we seek to establish.

\subsection{Bosonic string theory on \texorpdfstring{$\mathrm{AdS}_3$}{AdS3} and the WZW model}
We consider bosonic string theory on a background $\mathrm{AdS}_3 \times X$, where $X$ represents the internal space, that is described by some CFT that we also denote by $X$ and we do not need to specify further. We shall only require that $X$ is a unitary CFT with discrete spectrum and the central charge is
\be 
c_X=26-\frac{3k}{k-2}\ .
\ee
Since the central charge of the $\mathrm{SL}(2,\mathbb{R})_k$ model that describes $\mathrm{AdS}_3$ is $\frac{3k}{k-2}$, this ensures criticality of the background.\footnote{Strictly speaking, $\mathrm{AdS}_3$ string theory is described by the WZW model based on the universal cover of $\mathrm{SL}(2,\mathbb{R})$.}

This string theory suffers from the presence of a tachyon and genus $g \ge 1$ correlators are ill-defined. For this reason and for technical simplicity, we will only consider $g=0$ correlators here. We will consider the superstring in Section~\ref{sec:superstrings}.

\paragraph{The spectrum of string theory.} The spectrum was analyzed many times in the literature and so we only briefly recall it here. We follow the conventions of \cite{Maldacena:2000hw}. Representations of the affine algebra $\mathfrak{sl}(2,\mathbb{R})_k$ controlling the $\mathrm{SL}(2,\mathbb{R})_k$ WZW model are labeled by an $\mathfrak{sl}(2,\mathbb{R})$ spin $j$ and spectral flow $w \in \mathbb{Z}_{\ge 0}$.\footnote{One can restrict to $w \ge 0$ without loss of generality, since the $w<0$ are understood as the corresponding bra states in the dual CFT.} The ground states of the representation are also labeled by the eigenvalue of $J_0^3$, which in view of holography we denote by $h$, since gets mapped to the conformal weight of the state in the dual CFT. The affine primary ground states form a representation of spin $j$ under the spectrally flowed zero-mode algebra $J_w^+$, $J_0^3-\frac{kw}{2}$ and $J_{-w}^-$ of $\mathfrak{sl}(2,\mathbb{R})_k$.

The spin $j$ can take values in $\mathbb{R}$ or in $\frac{1}{2}+i \mathbb{R}$, corresponding to discrete series and principal continuous series representations. Normalizability of the state (or alternatively unitarity of string theory on the background) imposes the inequality $\frac{1}{2}<j<\frac{k-1}{2}$ for discrete representations. Discrete representations of type $[\mathcal{D}^+_j]^w$ have a highest weight state with $h=j+\frac{kw}{2}$ that is annihilated by all modes $J^-_m$ with $m \ge -w$, $J^3_m$ with $m>0$ and $J^+_m$ with $m >w$. Similarly, discrete representations of type $[\mathcal{D}^-]^w$ possess a highest weight state with $h=-j+\frac{kw}{2}$ that is annihilated by all mode $J^-_m$ with $m>-w$, $J^3_m$ with $m>0$ and $J^+_m$ with $m \ge w$. Due to the identification 
\be 
[\mathcal{D}^+_j]^w=[\mathcal{D}^-_{\frac{k}{2}-j}]^{w+1} \ , \label{eq:D+D- identification}
\ee
discrete representations of type $[\mathcal{D}^-_j]^w$ do not have to be included separately in the spectrum.
The model also includes continuous series representations with $j=\frac{1}{2}+i s$ and $s \in \mathbb{R}$. For these representations, the ground state representation is unbounded from below and above. These representations enjoy a reflection symmetry and representations with spins $j$ and $1-j$ are equivalent.

The quadratic Casimir of the spectrally flowed zero-mode algebra of all these representations is $\mathcal{C}(j)=-j(j-1)$. Consequently, the worldsheet conformal weight of these spectrally flowed affine primary states is
\be 
\Delta=-\frac{j(j-1)}{k-2}-w h+\frac{k}{4}w^2\ . \label{eq:Delta}
\ee
Including also a contribution $\Delta_X$ of the internal CFT to the worldsheet conformal weight, the mass-shell condition of string theory hence reads
\be 
\Delta+\Delta_X=-\frac{j(j-1)}{k-2}-w h+\frac{k}{4}w^2+\Delta_X\overset{!}{=}1\ . \label{eq:mass shell condition}
\ee
From this one determines the spacetime conformal weight in terms of the worldsheet conformal weight to be
\be 
h=-\frac{j(j-1)}{w(k-2)}+\frac{k}{4}w+\frac{\Delta_X-1}{w}\ .
\ee
It is useful to define the quantities
\be 
b\equiv \frac{1}{\sqrt{k-2}}\ , \qquad Q\equiv b^{-1}-b=\frac{k-3}{\sqrt{k-2}}\ ,\qquad \alpha\equiv \frac{j+\frac{k}{2}-2}{\sqrt{k-2}}\ . \label{eq:map of parameters}
\ee
We have 
\be 
\frac{Q}{2}=\frac{1}{2b}-\frac{b}{2}<\alpha<\frac{1}{b}-\frac{b}{2}
\ee
for normalizable discrete representations and $\alpha \in \frac{Q}{2}+i \mathbb{R}$ for principal continuous representations. The conformal weight can be written as
\be 
h=\frac{k(w^2-1)}{4w}+\frac{\Delta_X}{w}+\frac{\alpha(Q-\alpha)}{w}\ . \label{eq:mass shell condition solution h}
\ee
As we shall review in Section~\ref{subsec:symmetric product orbifold}, this agrees for the principal continuous representations with the spectrum of the symmetric orbifold theory
\be 
\text{Sym}^N(\mathbb{R}_Q \times X) \ ,\label{eq:sym orbifold}
\ee
where $\mathbb{R}_Q$ denotes the linear dilaton theory with slope $Q$. However, this cannot be the correct spacetime CFT, since for example in the linear dilaton theory, the states corresponding to $\alpha$ and $Q-\alpha$ are different (for $\alpha \in \frac{Q}{2}+i \mathbb{R}$), but they are identified by reflection symmetry on the worldsheet. We will see the correct statement below.

\paragraph{Three-point functions of the worldsheet theory.} The spectrum is not enough to identify the dual CFT and hence we will also study correlators, focusing mainly on three-point functions. There are vertex operators associated to the spectrally flowed affine primary states that we discussed above, whose $\mathrm{AdS}_3$-part will be denoted by
\be 
V_{j,h}^w(x;z)\ .
\ee
They can be multiplied with any vertex operator from the internal CFT.
They depend beyond the usual worldsheet coordinate $z$ also on a dual coordinate $x$, that gets identified with the coordinate in the dual CFT. It can formally be introduced by writing
\be 
V_{j,h}^w(x;z)=\mathrm{e}^{x J_0^+} V_{j,h}^w(0;z) \mathrm{e}^{-x J_0^+}\ , \label{eq:definition x coordinate}
\ee
since $J_0^+$ is the translation operator in the dual CFT space. The vertex operators $V_{j,h}^w(x;z)$ transform under two separate copies of M\"obius groups that act on $x$-space with weight $h$ and on $z$-space with weight $\Delta$ given by \eqref{eq:Delta}. There is also an anti-holomorphic label $\bar{h}$ for these vertex operators that we suppress in the following. By virtue of the definition \eqref{eq:definition x coordinate}, these vertex operators are coherent states from the standard CFT point of view and consequently enjoy some peculiar properties. We sometimes write $V_{j,h}^{\pm,w}(x;z)$ to emphasize that the vertex operators belong to the discrete representation $[\mathcal{D}_j^\pm]^w$.

In \cite{Dei:2021xgh}, a proposal for their three-point functions was made based on various consistency conditions. It reads for continuous spectrally flowed representations
\begin{multline}
\left\langle V^{w_1}_{j_1,h_1}(0; 0) \,  V^{w_2}_{j_2,h_2}(1;  1) \, V^{w_3}_{j_3,h_3}(\infty;\infty) \right\rangle =\int \prod_{i=1}^3 \frac{\mathrm{d}^2 y_i}{\pi}\ \prod_{i=1}^3 \left|y_i^{\frac{k w_i}{2}+j_i-h_i-1}\right|^2 \\
\times\begin{cases} D(j_1,j_2,j_3)\left|X_\emptyset^{\sum_l j_l-k}\displaystyle\prod_{i<\ell}^3  X_{i \ell}^{\sum_l j_l-2j_i-2j_\ell}\right|^2 \  , & \sum_i w_i \in 2\mathds{Z}\ ,\\
\mathcal N(j_1)D(\tfrac{k}{2}-j_1,j_2,j_3) \left|X_{123}^{\frac{k}{2}-\sum_l j_l} \displaystyle\prod_{i=1}^3 X_i^{\sum_l j_l-2j_i-\frac{k}{2}}\right|^2 \ ,  & \sum_i w_i \in 2\mathds{Z}+1\ .
\end{cases} \label{eq:3-point function worldsheet}
\end{multline}
Here, 
\be 
D(j_1,j_2,j_3)=-\frac{\G_k(1-j_1-j_2-j_3)}{2\pi^2 \nu^{j_1+j_2+j_3-1} \lgamma\left(\frac{k-1}{k-2}\right) }\prod_{i=1}^3 \frac{\G_k(2j_i-j_1-j_2-j_3)}{\G_k(1-2j_i)} \label{eq:unflowed three-point function}
\ee
is the three-point function of three unflowed vertex operators \cite{Teschner:1997ft}. It is given in terms of the Barnes double Gamma function  $\G_k(x)$ that satisfies the following functional identities:
\begin{subequations}
\begin{align} 
\G_k(j+1)&=\lgamma\left(-\frac{j+1}{k-2}\right)\G_k(j)\ , \\
\G_k(j-k+2)&=\frac{\lgamma(j+1)}{(k-2)^{2j+1}} \G_k(j)\ , \\
\G_k(j)&=\G_k(-j+1-k)\ ,
\end{align} \label{eq:Barnes double Gamma function functional identities}
\end{subequations}
where $\lgamma(x)=\Gamma(x)/\Gamma(1-x)$.
Furthermore,
\be 
\mathcal{N}(j)=\frac{\nu^{\frac{k}{2}-2j}}{\lgamma\left(\frac{2j-1}{k-2}\right)}
\ee
is a normalization factor related to the two-point function. Finally, $X_I$ for $I$ a subset of $\{1,2,3\}$ is defined as\footnote{This definition makes sense for $|I|=w_1+w_2+w_3 \bmod 2$, which is responsible for the case distinction in \eqref{eq:3-point function worldsheet}.}
\be 
X_I= \sum_{i \in I:\ \varepsilon_i=\pm 1} P_{(w_1+\varepsilon_1,w_2+\varepsilon_2,w_3+\varepsilon_3)}\prod_{i\in I} y_i^{\frac{1-\varepsilon_i}{2}} \ . 
\ee
The quantity $P_{(w_1,w_2,w_3)}$ that appears here is defined for $w_1+w_2+w_3$ even as
\be 
P_{(w_1,w_2,w_3)}=S_{(w_1,w_2,w_3)} G\left(\tfrac{w_1+w_2+w_3}{2} +1\right)  \prod_{i=1}^3 \frac{G(\frac{w_1+w_2+w_3}{2}-w_i+1)}{G(w_i+1)}\ ,
\ee
where $G(n)=\G_3(n)=\prod_{i=1}^{n-1} \Gamma(i)$ is the Barnes double factorial.
$S_{(w_1,w_2,w_3)}$ is a phase that can be written as
\be 
S_{(w_1,w_2,w_3)}=(-1)^{\frac{1}{2}x(x+1)}\ , \qquad x=\frac{1}{2} \sum_{i=1}^3 (-1)^{w_iw_{i+1}}w_i\ ,
\ee
where indices are understood to be modulo 3. Here and in the following, we also use the following abbreviation:
\be 
|x^a|^2 \equiv x^a \bar{x}^{\bar{a}}\ ,
\ee
where $\bar{a}$ is not the complex conjugate of $a$, but rather the corresponding right-moving quantity. E.g.\ in the exponent of $y_i$ in the formula for the 3-point function, $\bar{h}_i$ might differ from $h_i$ by an integer.
This explains all ingredients entering \eqref{eq:3-point function worldsheet}.

The three-point functions depend on a parameter $\nu$. This parameter can be viewed as a cosmological constant in the worldsheet theory. It is however not a parameter of the theory and can always be removed by a field redefinition. The only meaningful parameter in the string theory will be the combination $C_{\mathrm{S}^2} \, \nu$ of the coupling constant $\nu$ and the normalization of the sphere path integral that we will introduce below. This combination is identified with $g_\text{string}^{-2}$. We could set $\nu$ to any fixed value, for example \cite{Teschner:1999ug} chooses
\be 
\nu=\pi(k-2)\, \lgamma\left(\frac{k-3}{k-2} \right)\ ,
\ee
but we will leave $\nu$ unspecified.

The formula also applies to discrete representations in the following sense. For discrete representations of type $\mathcal{D}^+$, the $y_i$-exponent in the first line is in $\mathbb{Z}_{\le -1}$, which leads to a divergence of the integral from the region $y_i \sim 0$. Similarly for representations $\mathcal{D}^-$, there is a divergence from the region $y_i \sim \infty$. To extract the discrete representation correlator, one should extract the residues in the corresponding weight $h_i$. For example, if the first vertex operator is a discrete representation of type $\mathcal{D}^+_j$, the formula for the correlator becomes
\begin{multline}
\left\langle V^{+,w_1}_{j_1,h_1,\bar{h}_1}(0; 0) \,  V^{w_2}_{j_2,h_2,\bar{h}_2}(1;  1) \, V^{w_3}_{j_3,h_3,\bar{h}_3}(\infty;\infty) \right\rangle\\
=\Res_{\delta h_1=0}\left\langle V^{w_1}_{j_1,h_1+\delta h_1,\bar{h}_1+\delta h_1}(0; 0) \,  V^{w_2}_{j_2,h_2,\bar{h}_2}(1;  1) \, V^{w_3}_{j_3,h_3,\bar{h}_3}(\infty;\infty) \right\rangle\ ,
\end{multline}
where we also wrote the right-moving spacetime conformal weights for extra clarity and on the RHS we insert \eqref{eq:3-point function worldsheet}.

Finally, eq.~\eqref{eq:3-point function worldsheet} also applies to the unflowed vertex operators as follows. Because continuous series representations do not lead to physical states in string theory, we are only interested in discrete representations of type $\mathcal{D}^+_j$ in the unflowed sector. Hence we should use the residue prescription as explained above. From the definition of the $X_I$'s, one can check that the second line in \eqref{eq:3-point function worldsheet} becomes independent of $y_i$ if the $i$-th vertex operator is unflowed. So, one ends up with 
\be 
\Res_{\delta h_i=0} \int \frac{\mathrm{d}^2 y_i}{\pi} \ y_i^{j_i-h_i-\delta h_i-1}\bar{y}_i^{j_i-\bar{h}_i-\delta h_i-1} = \delta_{h_i,j_i} \delta_{\bar{h}_i,j_i}\ .
\ee
This means that $h_i=\bar{h}_i=j_i$ in the unflowed sector, which can be also seen directly from representation theory. Consequently, with this understanding one can simply drop the $y_i$-integral together with the explicit appearance of $y_i$ in the first line of \eqref{eq:3-point function worldsheet} if the $i$-th vertex operator comes from an unflowed discrete representation.

\paragraph{Worldsheet identity operator and the two-point function.} The discrete representation vertex operator $V_{j,h}^{+,w}(x;z)$ with $j=0$ and $w=0$ is the identity of the worldsheet theory (that is not part of the normalizable worldsheet spectrum). We normalize vertex operators on the worldsheet such that the insertion of this operator reduces a 3-point function to a 2-point function. Explicitly, we have for the worldsheet two-point function
\begin{align}
&\left\langle V^{w_1}_{j_1, h_1}(0;0) \,  V^{w_2}_{j_2, h_2}(\infty;\infty) \right\rangle\nonumber \\
&\qquad=\delta_{w_1,w_2} \int \prod_{i=1}^2 \frac{\mathrm{d}^2 y_i}{\pi} y_i^{\frac{kw_i}{2}-h_i+j_i-1}\bar{y}_i^{\frac{kw_i}{2}-\bar{h}_i+j_i-1}\ \Big(  B(j_1) \delta(j_1-j_2) \left|1- y_1 y_2 \right|^{-4j_1}
\nonumber\\
&\hspace{5cm}+\delta(j_1+j_2-1) |y_1|^{-2j_1}|y_2|^{-2j_2} \delta^{(2)} \left(1- y_1 y_2 \right)\Big) \\
&\qquad=4 i \,   \delta^{(2)}(h_1-h_2) \, \delta_{w_1,w_2} \Bigl( R(j_1,h_1,\bar h_1) \, \delta(j_1-j_2) + \delta(j_1+j_2-1) \Bigr) \ , \label{eq:2-point function worldsheet}
\end{align}
where
\begin{subequations}
\begin{align} 
B(j)&=\frac{k-2}{\pi} \frac{\nu^{1-2j}}{\lgamma\left(\frac{2j-1}{k-2}\right)}\ , \\
 R(j,h,\bar{h})&=  \frac{(k-2)\nu^{1-2j}\lgamma(h-\frac{kw}{2}+j)}{\lgamma\left(\frac{2j-1}{k-2}\right)\lgamma(h-\frac{kw}{2}+1-j)\lgamma(2j)}
\end{align}
\end{subequations}
are the normalization of the two-point function and the reflection coefficient and $\delta^{(2)}(h)=\delta(h+\bar{h})\delta_{h,\bar{h}}$.\footnote{In these formulas, we write $\lgamma(x)=\Gamma(x)/\Gamma(1-\bar{x})$. Here, $\bar{x}$ again represents the corresponding right-moving quantity. For example, $\lgamma(h-\frac{kw}{2}+j)$ appear in the numerator of $R(j,h,\bar{h})$ means $\Gamma(h-\frac{kw}{2}+j)/\Gamma(1-\bar{h}+\frac{kw}{2}-j)$.}

\paragraph{String theory two- and three-point functions.} While we have described the worldsheet two- and three-point functions, we still have to explain how to obtain the corresponding string theory correlators. We shall denote string correlators by double brackets. For the three-point function, we simply have
\be 
\langle\!\langle \mathcal{O}^{w_1}_{j_1}(0) \,\mathcal{O}^{w_2}_{j_2}(1)\, \mathcal{O}^{w_3}_{j_3}(\infty)  \rangle\!\rangle=C_{\mathrm{S}^2} \left\langle V^{w_1}_{j_1,h_1}(0; 0) \,  V^{w_2}_{j_2,h_2}(1;  1) \, V^{w_3}_{j_3,h_3}(\infty;\infty) \right\rangle\ . \label{eq:string three point function}
\ee
The weights $(h_i,\bar{h}_i)$ on the right-hand side are determined by the mass-shell condition \eqref{eq:mass shell condition solution h}. The constant $C_{\mathrm{S}^2}$ determines the normalization of the string path integral (see e.g.\ \cite{Polchinski:1998rq} for a detailed discussion of this normalization in flat space string theory). We will determine it from consistency later.
The operators $\mathcal{O}^{w_i}_{j_i}(x)$ are to be identified with operators in the dual CFT that we seek to find. Of course, we could also include non-trivial vertex operators of the internal CFT $X$. To avoid cluttering of the notation, we will just consider these simple vertex operators where the internal CFT is in the vacuum, but it will be obvious that everything works equally well for vertex operators that are excited in the internal space.

The string two-point function is more subtle, because of two competing effects. The mass-shell condition 
\be 
-\frac{j(j-1)}{k-2}-w h+\frac{k}{4}w^2=1
\ee
together with $h_1=h_2$ and $\bar{h}_1=\bar{h}_2$ implies that $j_1=j_2$ or $j_1=1-j_2$. In view of \eqref{eq:2-point function worldsheet} this means that the worldsheet two-point function becomes divergent on the mass-shell condition. However in string theory, we should also divide by the subgroup of the M\"obius transformation that fixes two-points. Since it has infinite volume, it should remove the divergence. This is very similar to what happens in flat space \cite{Erbin:2019uiz}. It was argued in \cite{Maldacena:2001km}, that the volume of the M\"obius symmetry cancels the delta functions $\delta(j_1-j_2)$ or $\delta(j_1+j_2-1)$, but leaving a finite ratio that was claimed to be $|2j_1-1|$. Actually our computation will give an independent derivation of this fact. Notice now that the mass-shell condition eq.~\eqref{eq:mass shell condition} implies
\begin{align}
\delta^{(2)}(h_1-h_2)&=\delta\left(-\frac{2(j_1-j_2)(j_1+j_2-1)}{w(k-2)}\right) \\
&=\frac{w(k-2)}{2|2j_1-1|} \left(\delta(j_1-j_2)+\delta(j_1+j_2-1)\right)\ .
\end{align}
Hence one can write the string theory two-point function as\footnote{The factor of $i$ is a bit subtle and we will not be too careful about it when comparing with the dual CFT.}
\be 
\langle\!\langle \mathcal{O}^{w_1}_{j_1}(0) \,\mathcal{O}^{w_2}_{j_2}(\infty) \rangle\!\rangle=2 iw(k-2) C_{S^2} \delta_{w_1,w_2} \Bigl( R(j_1,h_1,\bar h_1) \, \delta(j_1-j_2) + \delta(j_1+j_2-1) \Bigr) \ .
\ee

\subsection{The symmetric product orbifold and its deformation} \label{subsec:symmetric product orbifold}
On the other side of the duality, we consider a symmetric product orbifold
\be 
\text{Sym}^N(\mathbb{R}_Q \times X) \ ,
\ee
as we have already anticipated in \eqref{eq:sym orbifold}. Notice that the central charge of the seed theory $\mathbb{R}_Q \times X$ is
\be 
c=1+6 Q^2+c_X=1+\frac{6(k-3)^2}{k-2}+26-\frac{3k}{k-2}=6k\ .
\ee
The three-point functions of this theory do not agree with \eqref{eq:3-point function worldsheet} and hence among other reasons this theory cannot be the correct dual. We will momentarily describe the correct modification, but let us first review the computation of correlation functions in the symmetric orbifold theory. For much more details we refer to \cite{Dixon:1986qv, Hamidi:1986vh, Arutyunov:1997gt, Arutyunov:1997gi, Jevicki:1998bm, Lunin:2000yv, Lunin:2001pw, Pakman:2009zz, Roumpedakis:2018tdb, Dei:2019iym}.

\paragraph{Twist fields.} We will focus on single-twist correlation functions, since these map holographically to single string correlators. A twist operator is labeled by the length of the twist $w\in \mathbb{Z}_{\ge 1}$ and has conformal weight $\frac{c(w^2-1)}{24w}$, where $c=6k$ is the central charge of the seed theory. For every vertex operator with weights $(h,\bar{h})$ with $h-\bar{h} \in w \mathbb{Z}$ in the seed theory, there is a corresponding vertex operator in the twisted sector $w$ with weight
\be 
h_w=\frac{c(w^2-1)}{24w}+\frac{h}{w}
\ee
and similarly for $\bar{h}$.
For the the vacuum, the corresponding vertex operator corresponds to the twisted sector ground state. Otherwise,
we sometimes call these twist fields `dressed'. In the present case, we can give twist fields momenta in the linear dilaton direction, which leads to twist fields with conformal dimension
\be 
h=\bar{h}=\frac{k(w^2-1)}{4w}+\frac{\alpha(Q-\alpha)}{w}\ . \label{eq:conformal weight dressed twist field}
\ee
As noted earlier, this agrees with the relation coming from the worldsheet \eqref{eq:mass shell condition solution h}. We will denote these dressed twist fields by $\sigma_{w,\alpha}(x)$.

\paragraph{Correlators of the symmetric product orbifold.} Correlators of (dressed) twist fields can be computed in two steps. The full twist operator is given by summing over elements in the conjugacy class of the permutation $(1 \cdots w)$ as follows,
\be 
\sigma_w(x)=\frac{\sqrt{(N-w)!w}}{\sqrt{N!}} \sum_{\tau \in [(1 \cdots w)] \subset S_N} \sigma_\tau(x)\ .
\ee
This reduces the problem of computing correlators of twist operators $\sigma_w(x)$ to computing correlators of `gauge-dependent' twist operators $\sigma_\tau(x)$. 

Correlation functions of gauge-dependent twist operators
\be 
\left\langle \prod_{i=1}^n \sigma_{\tau_i}(x_i) \right \rangle
\ee
are then computed as follows. In these correlation functions, some fixed order is chosen because of the presence of branch cuts (which corresponds to the `gauge choice'). By including the appropriate combinatorial factor, we can assume that $(\tau_1,\dots,\tau_n)$ act only on the copies $1,\dots,d$ for some $d$ (and consider $(\tau_1,\dots,\tau_n)$ that arise from relabeling the sites as equivalent). The correlator is only non-vanishing when $\prod_{i=1}^n \tau_i=\mathds{1}$, where the product is taken in the order that we fixed earlier. This data precisely determines a branched covering surface whose sheets are labeled by the numbers $1,\dots,d$ and the monodromies around the branch points realize the group elements $\tau_i$. This surface can have higher genus (and even be disconnected); its genus $g$ can be determined from $d$ and the twists by the Riemann-Hurwitz formula
\be 
2-2g=2d-\sum_i (w_i-1)\ . \label{eq:Riemann-Hurwitz formula}
\ee
By definition there is a branched covering map from the covering surface to the sphere parametrized by $x_i$. The covering map $\gamma(\zeta)$ has by construction the following Taylor expansion near the ramification points $z_i$ in the covering surface:
\be 
\gamma(\zeta)=x_i+a_i(\zeta-z_i)^{w_i}+\dots
\ee
The covering map determines the quantities $a_i$, which play an important role below. The covering map also has poles, near which one can compute the residues
\be 
\gamma(\zeta) \sim \frac{\Pi_a}{\zeta-z_a}+\mathcal{O}(1)\ .
\ee
The product of the residues $\Pi \equiv \prod_a \Pi_a$ also enters in the formulas below.\footnote{We always put the first field at $x=0$, the second at $x=1$ and the third field at $x=\infty$. The presence of a field at $x=\infty$ leads to some special features. First, the $a_i$ near $\infty$ is defined as
\be 
\gamma(\zeta)=\frac{(-1)^{w_3+1}}{a_3 \zeta^{w_3}}+\mathcal{O}(\zeta^{-w_3-1})\ .
\ee
Since the covering map has also less poles than generically (some poles have been swallowed by the third field), there is also a correction factor for $\Pi$ needed to arrive at symmetrical formulas:
\be 
\Pi \equiv w_3^{-w_3-1} \prod_a \Pi_a\ .
\ee 
}

One can then compute the correlator by lifting it to the covering space. Since the induced metric on the covering space is not flat, there is a Weyl factor that accounts for the non-trivial metric. If the covering surface has genus 0, it can explicitly be computed in terms of the quantities $a_i$ and $\Pi$, which leads to the result
\be 
\left\langle \prod_{i=1}^n \sigma_{\tau_i,\alpha_i}(x_i) \right \rangle=\left| \prod_{i=1}^n w_i^{-\frac{k(w_i+1)}{4}} a_i^{\frac{k(w_i-1)}{4}-h_i} \Pi^{-\frac{k}{2}}\right|^2 \left\langle \prod_{i=1}^n \mathcal{O}_{\alpha_i}(z_i) \right \rangle\ .
\ee
Here the quantities $a_i$ and $\Pi$ are those that arise from the covering map that is specified by the $\tau_i$'s. 
The operators $\mathcal{O}_{\alpha_i}$ are the untwisted vertex operators with momentum $\alpha_i$. A similar formula should hold for higher genus covering surfaces, but the prefactor from the Weyl rescaling does not enjoy such an explicit formula.
The correlator on the RHS has to be evaluated on the corresponding covering surface (with the flat metric in the genus 0 case).

Finally, correlators of gauge-invariant twist operators are given by
\begin{multline} 
\left\langle \prod_{i=1}^n \sigma_{w_i,\alpha_i}(x_i) \right \rangle=\frac{N!}{(N-d)!}\prod_{i=1}^n \frac{\sqrt{(N-w_i)!w_i}}{N!} \\ \times \sum_{\text{covering surfaces}}\left| \prod_{i=1}^n w_i^{-\frac{k(w_i+1)}{4}} a_i^{\frac{k(w_i-1)}{4}-h_i} \Pi^{-\frac{k}{2}}\right|^2 \left\langle \prod_{i=1}^n \mathcal{O}_{\alpha_i}(z_i) \right \rangle\ , \label{eq:correlator symmetric orbifold}
\end{multline}
where again the simple prefactor in the second line only applies to the case of genus 0 covering surfaces.
The combinatorial prefactor can be expanded for large $N$, which leads to
\be 
\frac{N!}{(N-d)!}\prod_{i=1}^n \frac{\sqrt{(N-w_i)!w_i}}{\sqrt{N!}} \sim \prod_{i=1}^n \sqrt{w_i} N^{1-g-\frac{n}{2}}=\prod_{i=1}^n \sqrt{w_i} N^{\frac{\chi}{2}}\ , \label{eq:N scaling}
\ee
where we used the Riemann-Hurwitz formula \eqref{eq:Riemann-Hurwitz formula} and $\chi$ is the Euler characteristic of the covering surface. 

We will focus on the leading large $N$ behavior. As we mentioned, the covering space could be connected or disconnected. The leading connected contribution comes from genus 0 covering spaces, for which we can also write explicitly
\be 
\left\langle \prod_{i=1}^n \mathcal{O}_{\alpha_i}(z_i) \right \rangle=b\, \delta\left(\sum_i \alpha_i-Q \right) \left|\prod_{i<j} (z_i-z_j)^{-2\alpha_i\alpha_j} \right|^2\ , \label{eq:genus 0 linear dilaton correlator}
\ee
where the delta-function imposes anomalous momentum conservation. The factor of $b=(k-2)^{-\frac{1}{2}}$ is a convention that makes the comparison with the worldsheet analysis simpler.

\paragraph{Marginal deformation.} The theory has a non-normalizable marginal operator given by
\be 
\Phi(x)\equiv \sigma_{2,\alpha=-\frac{1}{2b}}(x)\ ,
\ee
where we recall from \eqref{eq:map of parameters} that $b=(k-2)^{-\frac{1}{2}}$. It is straightforward to check from eq.~\eqref{eq:conformal weight dressed twist field} that 
\be 
h=\frac{k(2^2-1)}{4 \cdot 2}-\frac{1}{2} \cdot\frac{1}{2b}\left(Q+\frac{1}{2b}\right)=1\ ,
\ee
where we also recall that $Q=b^{-1}-b$. Since the two-point function in the deformed theory that we compute in Section~\ref{subsec:two-point function} does receive corrections, the operator is also exactly marginal.
This operator creates an exponential wall and is hence in many ways very similar to the exponential operator in Liouville theory. Let us denote the linear dilaton direction by $\phi$, so that the corresponding untwisted vertex operator would be $\mathrm{e}^{-\frac{\phi}{b}}$. 

Delta-function normalizable states of the theory are plane waves that scatter off the exponential wall. Since $b>0$, the exponential wall becomes weak at $\phi \to \infty$. Hence in this regime the spectrum remains unchanged and thus the scattering states of the theory have identical spectrum to the undeformed theory, except that the states will be superpositions of incoming and outgoing waves. We will explain the appearance of normalizable bound states in Section~\ref{sec:poles and short strings}.

\paragraph{KPZ scaling.}
Let us denote the coupling constant of the interaction as $\mu$.  The dependence of correlation functions of $\mu$ follows from a slight modification of the usual KPZ scaling argument \cite{Knizhnik:1988ak}. At leading order in large $N$, the covering space is necessarily a sphere as follows from \eqref{eq:N scaling}. One can then apply the KPZ scaling argument to conclude that
\be 
\left\langle \prod_{i=1}^n \sigma_{w_i,\alpha_i}(x_i) \right \rangle_{\!\!\mu} \propto \mu^{2b(\sum_i \alpha_i-Q)}\  \label{eq:KPZ scaling}
\ee
to leading order in $N^{-1}$. We will discuss subleading orders in Section~\ref{subsec:large N}.
Since the exponent of $\mu$ is in general not a positive integer and the perturbing field is a non-normalizable vertex operator,  conformal perturbation theory is not well-behaved. However, as in Liouville theory, one can still compute correlation functions for which the $\mu$-exponent is a positive integer in conformal perturbation theory \cite{Zamolodchikov:1995aa}. These correlation functions exhibit a `bulk pole' so that one is actually computing the residue of the pole.\footnote{More precisely, one is computing $\pi$ times the residue. This is because of the Sokhotski-Plemelj formula
\be 
\lim_{\varepsilon \to 0^+}\frac{1}{x+i \varepsilon}=-i \pi \delta(x)+\mathcal{P}(x^{-1})\ ,
\ee
where $\mathcal{P}$ is the Cauchy principal value. The various factors of $i$ are somewhat subtle and we will not precisely keep track of them.}
 Assuming that the correlation function in the perturbed theory can be defined and is analytic as a function of $\alpha_i$, it has a pole at 
\be 
 2b\left(\sum_i \alpha_i-Q\right)=m \in \mathbb{Z}_{\ge 0}
\ee
with residue
\be 
\Res_{2b(\sum_i \alpha_i-Q)=m} \left\langle \prod_{i=1}^n \sigma_{w_i,\alpha_i}(x_i) \right \rangle_{\!\!\mu} 
=\frac{(-\mu)^m}{\pi m!} \int \prod_{\ell=1}^m \mathrm{d}^2\xi_\ell\ \left\langle \prod_{i=1}^n \sigma_{w_i,\alpha_i}(x_i)\prod_{\ell=1}^m \Phi(\xi_\ell) \right \rangle_{\!\! 0}'\ , \label{eq:residue conformal perturbation theory}
\ee
where the prime means that the momentum conserving delta-function is removed.\footnote{In order to match with the worldsheet conventions, we will take the residue in the spins $j_i$, which are related to the momenta $\alpha_i$ via \eqref{eq:map of parameters}. The prime then omits the delta-function and the constant $b$ from the correlator \eqref{eq:genus 0 linear dilaton correlator}.} Essentially, the delta-function of the undeformed theory is expected to become a pole in the interacting theory. 

We should mention that this deformation of the symmetric orbifold is technically simpler than the usual marginal deformation that is considered for example in the symmetric orbifold of $\mathbb{T}^4$. The reason is that the integrals in \eqref{eq:residue conformal perturbation theory} can be defined by analytic continuation in the external spins, whereas the integrals of the conformal perturbation theory of $\mathbb{T}^4$ have to be carefully regulated. Moreover, because of momentum conservation, there is essentially only one order in conformal perturbation theory that can contribute to a perturbed correlation function. See e.g.\ \cite{Benjamin:2021zkn} and references therein for a recent overview of the perturbation theory of $\text{Sym}^N(\mathbb{T}^4)$. 

We can use the explicit formula for symmetric orbifold correlators \eqref{eq:correlator symmetric orbifold} to compute the RHS explicitly. We will again focus on the leading large $N$ contribution. Moreover, we focus on the connected correlator on the left-hand side. This implies that also the correlator on the right-hand side has to be connected. Indeed, the dressed twist operators $\sigma_{w_i,\alpha_i}$ lift by assumption to the same component of the covering space and since we are considering the leading large $N$ contribution, this component should be a sphere. Momentum conservation holds in every component separately, and this forces all marginal operators to lie in the same component of the covering space. Hence any other component would not have any vertex operators and so should cover the sphere without ramifications, thus implying that any other component should also be a sphere. However, momentum conservation is violated for the sphere partition function without vertex operators and thus we conclude that the covering map has to be connected.\footnote{This is true in conformal perturbation theory, but beyond that we cannot make use of momentum conservation and the story is subtler. We return to it in Section~\ref{subsec:large N}.} We hence have from eqs.~\eqref{eq:correlator symmetric orbifold}, \eqref{eq:genus 0 linear dilaton correlator} and \eqref{eq:residue conformal perturbation theory}
\begin{multline}
\Res_{2b(\sum_i \alpha_i-Q)=m} \left\langle \prod_{i=1}^n \sigma_{w_i,\alpha_i}(x_i) \right \rangle_{\!\!\mu} =\frac{(-\sqrt{2}\mu)^m\prod_{i=1}^n \sqrt{w_i} }{\pi m! N^{\frac{n}{2}-1}} \\ \times\sum_{\begin{subarray}{c} \text{connected} \\ \text{covering surfaces}\end{subarray}}\int \prod_{\ell=1}^m \mathrm{d}^2\xi_\ell\ \Bigg| \, 2^{-\frac{3km}{4}}\prod_{i=1}^n w_i^{-\frac{k(w_i+1)}{4}} a_i^{\frac{k(w_i-1)}{4}-h_i} \prod_{i=n+1}^{m+n} a_i^{\frac{k}{4}-1} \Pi^{-\frac{k}{2}}  \\
\times\prod_{1\le i<j\le n} (z_i-z_j)^{-2\alpha_i\alpha_j}\prod_{i=1}^n \prod_{j=1}^m (z_i-\zeta_j)^{\frac{\alpha_i}{b}} \prod_{1\le i\le j \le m} (\zeta_i-\zeta_j)^{-\frac{1}{2b^2}} \Bigg|^2\ . \label{eq:deformed correlation function}
\end{multline}
We denoted insertion points for the marginal operators by greek letters to facilitate notation, i.e. $\xi_i$ and $\zeta_i$ denote the branch points in the base space and the covering space that are introduced by the presence of the marginal operators.
Our goal will be to compute the right hand side for two- and three-point function explicitly and show that it agrees with the worldsheet theory. 

\paragraph{Parity.} We should mention that there is a qualitative difference depending on the parity of $\sum_i (w_i-1)$ of the vertex operators involved in the correlator. The Riemann-Hurwitz formula \eqref{eq:Riemann-Hurwitz formula} implies that a covering map only exists when $\sum_i(w_i-1)$ is even. However, every insertion of a marginal operator changes the parity of $\sum_i(w_i-1)$. This means that for $\sum_i(w_i-1)$ even, only even orders in conformal perturbation theory can be non-zero, whereas for $\sum_i (w_i-1)$ odd, only odd orders can contribute. As a consequence, the result for the correlators looks significantly different depending on the parity of $\sum_i(w_i-1)$, just as it does in the string correlators. This can be seen explicitly from the formula for the string three-point functions in eq.~\eqref{eq:3-point function worldsheet}. 

\paragraph{Path-integral perspective.} From a path-integral perspective, one can see that the correlators will have poles. The action can be written as
\begin{multline} 
S=\sum_i \left[\frac{1}{4\pi}\int \mathrm{d}^2x\ \left(4\partial \phi^{(i)} \bar{\partial} \phi^{(i)}+Q R \phi^{(i)}\right) +S_X^{(i)}\Big|_\text{gauged}\right]+\mu \int \mathrm{d}^2 x \ \sigma_{2,\alpha=-\frac{1}{2b}}(x)\\
-2\sum_{i=1}^n \int \mathrm{d}^2 x\ \alpha_i \phi(x) \delta^2(x-x_i) \label{eq:deformed action}
\end{multline}
where `gauged' means that the $S_N$ action should be gauged in order to pass to the symmetric orbifold and we included some untwisted vertex operators in the second line. The following discussion is unchanged if these vertex operators would be twisted. We have $\phi(x)=\sum_i \phi^{(i)}(x)$.
As $\phi\to \infty$, the potential becomes weak and only the linear dilaton term and the vertex operators contribute. Thus the action becomes
\be 
S \sim 2\left(Q-\sum_i \alpha_i\right)\phi_\infty\ ,
\ee
where we used that the Gauss-Bonnet theorem on the sphere and $\phi_\infty$ denotes the zero-mode of $\phi$. From a path-integral perspective, we should require that
\be 
\Re\left(Q-\sum_i \alpha_i\right)>0 \label{eq:bound pole region}
\ee
for the path-integral to be well-defined. In particular, this region of the correlators should be free of singularities. For $Q-\sum_i \alpha_i=0$, there is a pole in the correlators. This is the first pole that is computed in 0th order conformal perturbation theory as discussed above.

\paragraph{$k>3$ vs.\ $k<3$.} From a path integral perspective, one can see a qualitative difference between the dual CFT depending on the sign of $k-3$ and hence $Q=\frac{k-3}{\sqrt{k-2}}$. This mirrors essentially what happens for the $\mathrm{AdS}_3$ string \cite{Giveon:2005mi}. For the continuum of the theory, $\alpha \in \frac{Q}{2}+i \mathbb{R}$ and for $n \ge 3$ point functions
\be 
\Re\left(Q-\sum_i \alpha_i\right)=-\frac{n-2}{2} Q\ ,
\ee
which violates the bound \eqref{eq:bound pole region} for $k>3$. This means that correlators of the delta-function normalizable states are only well-defined for $k<3$ from a path-integral perspective. For $k<3$, correlation functions of delta-function normalizable operators are always free of poles and can be found directly from the path integral. In that sense, the CFT that we are proposing is better-defined for $k<3$. However, it will turn out that the answer for correlators (at least for two- and three-point functions) is analytic in the momenta $\alpha_i$. Thus one can always compute them in a region where \eqref{eq:bound pole region} is satisfied and then analytically continue. Hence correlators for $k>3$ are uniquely defined. The spectrum of normalizable operators is however qualitatively different depending on the sign of $k-3$.

One can take this as an indication that the duality should only work for $k<3$ ($k<1$ in the supersymmetric case), as was assumed in \cite{Balthazar:2021xeh, Martinec:2021vpk}. We don't see any fundamental difficulty to define the theory CFT we are proposing also for $k>3$ (at least perturbatively in $N^{-1}$). Correlation functions can be defined for any $k$ by analytic continuation in $\alpha_i$ or directly in conformal perturbation theory.
While we discussed this for the genus 0 correlator (i.e.\ leading contribution in $N^{-1}$), the discussion applies equally well to higher genus contributions.

It might seem to the reader that the situation is similar to spacelike Liouville theory vs.\ timelike Liouville theory, that was discussed extensively in the literature \cite{Harlow:2011ny, Ribault:2015sxa}, but that is not the case. In the present case, $Q=b^{-1}-b$ (vs.\ $Q=b^{-1}+b$ in Liouville theory).
In the analytic continuation of Liouville theory, $b$ has to leave the real axis in order to leave the region $Q \ge 2$, where the path integral definition makes sense. However, because of the opposite sign, this is not the case here and there do not seem to be subtleties associated to the analytic continuation from $2<k<3$ to $k>3$. 

\section{Matching of two- and three-point functions} \label{sec:matching}
In this section, we will match the worldsheet two- and three-point functions of the bulk theory explicitly to the two- and three-point functions computed in conformal perturbation theory in the deformed symmetric orbifold theory. We will do so up to 4th order in perturbation theory. Concretely, this amounts to the equality
\begin{multline}
\Res_{\sum_i j_i=3-\frac{k}{2}+m\left(\frac{k}{2}-1\right)}\langle\!\langle \mathcal{O}^{w_1}_{j_1}(0) \,\mathcal{O}^{w_2}_{j_2}(1)\, \mathcal{O}^{w_3}_{j_3}(\infty)  \rangle\!\rangle\\
\overset{!}{=} \prod_{i=1}^3 N(w_i,j_i) \Res_{2b(\sum_i \alpha_i-Q)=m} \left\langle \prod_{i=1}^3 \sigma_{w_i,\alpha_i}(x_i) \right \rangle_{\!\!\mu}
\end{multline}
for $m \in \{0,1,2,3,4\}$ (and similarly for the two-point function). The condition on $\alpha$ on the RHS is the same as the condition on $j$ on the LHS via the identification \eqref{eq:map of parameters}. The left-hand side is given by \eqref{eq:string three point function} and the right hand side by \eqref{eq:deformed correlation function}. There are three undetermined quantities: the relative normalization of the vertex operators, which is accounted for the presence of the normalization factors $N(w_i,j_i)$, the normalization of the string path integral $C_{\mathrm{S}^2}$ and the relation between the coupling constants $\mu$ and $\nu$ in the CFT and string theory. We remind the reader that both $\mu$ and $\nu$ are not really parameters of the theory, but just set various normalizations. We will determine these constants by comparison. They turn out to be
\begin{subequations}
\begin{align}
C_{\mathrm{S}^2}&= \frac{N\nu^{k-4}}{32\pi^2 (k-2)^7\lgamma\left(\frac{k-1}{k-2}\right)^2}\ ,\\
N(w,j)&=  \frac{\sqrt{N}\nu^{\frac{k}{2}-2}w^{\frac{3}{2}-2j}}{4\pi (k-2)^3\lgamma\left(\frac{k-1}{k-2}\right)}\ ,\\
\mu&= -\frac{(k-2)2^{2k-\frac{5}{2}}\nu^{1-\frac{k}{2}}}{\pi}\ .
\end{align} \label{eq:matching of normalization constants}
\end{subequations}
Equality between the two sides amounts to highly non-trivial relations for the quantities $a_i$ and $\Pi$ for the covering map. While we do not have a proof for these relations, we check them numerically for a large set of choices for $w_i$.

\subsection{Qualitative matching}
Before launching into the computation in earnest, let us first explain how the matching works qualitatively. We will concentrate on the three-point function, since the two-point function can be obtained once we know the three-point function. The string correlator is expressed as a triple integral of the $y$-variables, see eq.~\eqref{eq:3-point function worldsheet}, whereas the CFT correlator at $m$-th order is expressed as an $m$-fold integral, see eq.~\eqref{eq:deformed correlation function}. As we have already explained the result depends on the parity of $m$, since the choices of spectral flows/twists obey the selection rule
\be 
\sum_i(w_i-1)=m \bmod 2\ .
\ee
Both on the string-side and on the CFT-side, the dependence on the conformal weights (that are in turn determined via \eqref{eq:mass shell condition solution h}) is very simple and implies that one should identify the variables $y_i$ with the covering map quantities $a_i$ on the CFT side. Thus, it will be advantageous to change integration variables from $\xi_\ell$ on the CFT side to $a_i$. 

At third order conformal perturbation theory, the number of integrals is the same and performing this change of variables will transform the CFT computation to the string answer. For $m<3$, there are too many integrals on the string side. In this case, the prefactor $D(j_1,j_2,j_3)$ or $\mathcal{N}(j_1)D(\frac{k}{2}-j_1,j_2,j_3)$ is regular for $\sum_i j_i=3-\frac{k}{2}+m\left(\frac{k}{2}-1\right)$ and the pole comes entirely from the $y$-integral as we will explain below. This allows us to perform some of the $y$-integrals, which will then match the CFT answer. In particular, the matching of these correlators works because of the identities (among others)
\begin{subequations}
\begin{align} 
X_i(a_1,a_2,a_3)&=0 \ , \qquad m=0\ , \\
X_{ij}(a_1,a_2,a_3)&=0 \ , \qquad m=1\ , \\
X_{123}(a_1,a_2,a_3)&=0 \ , \qquad m=2\ .
\end{align}
\end{subequations}
By this we mean e.g.\ that $a_1$, $a_2$ and $a_3$ of every covering map with ramification indices $(w_1,w_2,w_3,2,2)$ satisfy
\be 
X_{123}(a_1,a_2,a_3)=0\ .
\ee
This gives a geometric interpretation to the quantities $X_I$ that appear in the worldsheet answer \eqref{eq:3-point function worldsheet}. From the string side, their appearance was rather mysterious.

 For $m \ge 3$, the $y$-integral is non-singular and the pole on the worldsheet instead comes from the prefactors $D(j_1,j_2,j_3)$ or $\mathcal{N}(j_1)D(\frac{k}{2}-j_1,j_2,j_3)$. Hence in this case the string worldsheet representation of the answer is `simpler'. The main work is to show that the additional integrals in the CFT representation of the answer can be performed explicitly, so that we only end up with the three integrals over $a_1$, $a_2$ and $a_3$ in the end. 
 
From this explanation, it should be clear why we felt the need to check agreement of the string three-point functions with the CFT three-point functions up to 4th order. In the next two subsections, we explain the string and CFT computation in turn. These computations are somewhat technical and the reader is invited to skip them, since they are not needed for the further understanding of this paper.

We should also mention that the matching of two- and three-point functions does not require the use of the mass-shell condition, i.e.\ also off-shell correlators match. This is to be expected, since one could change the mass-shell condition by dressing the vertex operators under consideration by a vertex operator from the internal CFT $X$. Thus, the fact that the matching works off-shell implies that all these dressed correlators match as well.\footnote{We are however still not considering the most general two- and three-point functions, since one might also consider descendants in the $\mathrm{SL}(2,\mathbb{R})_k$ WZW model.}

\subsection{String computation} \label{subsec:string computation}
On the string side, our task is to compute the residue in the three-point function explicitly. We first analyze the prefactor $D(j_1,j_2,j_3)$ ($m$ odd) and $\mathcal{N}(j_1) D(\frac{k}{2}-j_1,j_2,j_3)$ ($m$ even) for $j_1+j_2+j_3=3-\frac{k}{2}+m\left(\frac{k}{2}-1\right)+\epsilon$. Let us write
\be 
D_m(j_1,j_2,j_3)=\begin{cases}
D(j_1,j_2,j_3) \ , & m\text{ odd}\ , \\
\mathcal{N}(j_1)D(\tfrac{k}{2}-j_1,j_2,j_3) \ , & m\text{ even}\ .
\end{cases}
\ee
 Using the function equations of the Barnes double Gamma function $\G_k(x)$, we find
 \begin{subequations}
\begin{align}
D_0(j_1,j_2,j_3)&=-\frac{\nu^{\frac{k}{2}-2}\prod_{i=1}^3 \lgamma(2-2j_i)}{2\pi^2(k-2)^2 \lgamma\left(\frac{k-1}{k-2}\right)\lgamma(k-2)} \ , \\
D_1(j_1,j_2,j_3)&=-\frac{\prod_{i=1}^3 \lgamma(2-2j_i)}{2\pi^2 \nu (k-2)\lgamma\left(\frac{k-1}{k-2}\right)}\ , \\
D_2(j_1,j_2,j_3)&=-\frac{\nu^{-\frac{k}{2}}}{2\pi^2 \lgamma\left(\frac{k-1}{k-2}\right)}\ , \\
D_m(j_1,j_2,j_3)&=\frac{\nu^{\frac{1}{2}(m-1)(2-k)-1}(k-2)^{m-2}\prod_{s=1}^{\lfloor \frac{m-3}{2} \rfloor}\lgamma\left(s(2-k)\right)}{2\pi^2\epsilon \lgamma\left(\frac{k-1}{k-2}\right)\prod_{i=1}^3 \prod_{s=1}^{\lceil \frac{m-3}{2} \rceil} \lgamma\left(2+(k-2)s-2j_i\right)}\ , & m&\ge 3\ . \label{eq:D prefactors}
\end{align} 
\end{subequations}
Because of the presence of $\frac{1}{\epsilon}$, the prefactor hence accounts for the singularity for $m \ge 3$. Thus, the residue for $m \ge 3$ on the string side is simply given by inserting this formula into \eqref{eq:3-point function worldsheet} and removing the $\epsilon^{-1}$.

We will now discuss the exceptional $m<3$ cases. Since they get harder as we decrease $m$, we will work backwards. 
\paragraph{$m=2$.} Notice that the exponent of $X_{123}$ becomes
\be 
\left|X_{123}^{\frac{k}{2}-j_1-j_2-j_3}\right|^2=\left|X_{123}^{-1-\epsilon}\right|\ .
\ee
Since the exponent for $\epsilon \to 0$ becomes $-1$, the integral diverges. We can use the formula
\be 
\Res_{\epsilon=0} \left|x^{-1-\epsilon}\right|=-\pi \delta^2(x)\ .
\ee
Consequently, the $y$-integral localizes onto the locus $X_{123}(y_1,y_2,y_3)=0$ after taking the residue. We also notice the identity\footnote{This identity and similar ones to follow can be checked numerically. The reader can find these checks in the ancillary \texttt{Mathematica} notebook.
We also suppress inconsequential signs that play no role in our computation.}
\be 
\partial_{y_3} X_{123}=\frac{X_1X_2}{X_3}\ ,
\ee
valid on the locus $X_{123}=0$.
This allows us to integrate out $y_3$, which leads to
\be 
C_{\mathrm{S}^2} \frac{\nu^{-\frac{k}{2}}}{2\pi^2 \lgamma\left(\frac{k-1}{k-2}\right)}\int \frac{\mathrm{d}^2 y_1 \ \mathrm{d}^2 y_2}{\pi^2} \left|\prod_{i=1}^3 y_i^{\frac{kw_i}{2}+j_i-h_i-1} \prod_{i=1}^2 X_i^{-2j_i} X_3^{2-2j_3} \right|^2 \label{eq:m=2 string result}
\ee
for the residue of the string correlator. In this formula, $y_3$, is implicitly determined through the condition $X_{123}(y_1,y_2,y_3)=0$.

\paragraph{$m=1$.} For $m=1$, the sum of the exponents of $X_{ij}$ is
\be 
-j_1-j_2-j_3=-2-\varepsilon\ .
\ee
This is a negative integer and hence leads to a divergence in the integral over $y_i$. The integral localizes now on the locus $X_{ij}=0$. The three conditions $X_{12}(y_1,y_2,y_3)=X_{13}(y_1,y_2,y_3)=X_{23}(y_1,y_2,y_3)=0$ are not independent and consequently this locus is one-dimensional, which will leave one integral. To integrate out $y_2$ and $y_3$, it is convenient to change variables to
\be 
y_2=y_2^0+\frac{ru}{\partial_{y_2} X_{23}}\ , \qquad y_3=y_3^0+\frac{r(1-u)}{\partial_{y_3} X_{23}}\ ,
\ee
where $y_2^0$ and $y_3^0$ is determined in terms of $y_1$ through the conditions $X_{ij}(y_1,y_2^0,y_3^0)=0$.
After change of variables, the $y_i$-integral becomes
\begin{multline}
\int \frac{\mathrm{d}^2 y_1\ \mathrm{d}^2 r\ \mathrm{d}^2 u}{\pi^3} \Bigg| \prod_{i=1}^3 y_i^{\frac{kw_i}{2}+j_i-h_i-1} X_\emptyset^{2-k}  r^{-1-\epsilon} u^{2j_3-2}(1-u)^{2j_2-2} \\
\times \frac{(\partial_{y_2} X_{12})^{2j_3-2}(\partial_{y_3} X_{13})^{2j_2-2}}{(\partial_{y_2}X_{23})^{2j_3-1}(\partial_{y_3}X_{23})^{2j_2-1}}\Bigg|^2\ .
\end{multline}
One can then extract the residue and simplify the answer further using the identity (valid on the locus $X_{ij}(y_1,y_2,y_3)=0$)
\be 
\partial_{y_2} X_{23} \partial_{y_3} X_{23}=X_\emptyset^2\ ,
\ee
which leads to the following answer for the string three-point function:
\begin{align}
\frac{C_{\mathrm{S}^2}}{2\pi^2 \nu (k-2)\lgamma\left(\frac{k-1}{k-2}\right)}\int \frac{\mathrm{d}^2 y_1}{\pi} \left| \prod_{i=1}^3 y_i^{\frac{kw_i}{2}+j_i-h_i-1} X_\emptyset^{-k} \left(\frac{\partial_{y_2} X_{12}}{\partial_{y_2} X_{23}}\right)^{2j_3-2}\left(\frac{\partial_{y_3} X_{13}}{\partial_{y_3} X_{23}}\right)^{2j_2-2}\right|^2\ . \label{eq:m=1 string result}
\end{align}
\paragraph{$m=0$.} In this case, the integral localizes on a single point $X_i(y_1,y_2,y_3)=0$ (which also implies $X_{123}(y_1,y_2,y_3)=0$). This locus can actually easily be described as $y_i=a_i$, where $a_i$ are the Taylor coefficients of the unique covering map that appears in the three-point function case. Explicitly,
\be 
a_i=\frac{\begin{pmatrix} \frac{1}{2} (w_i+w_{i+1}+w_{i+2}-1) \\\frac{1}{2} (-w_i+w_{i+1}+w_{i+2}-1) \end{pmatrix}}{\begin{pmatrix} \frac{1}{2} (-w_i+w_{i+1}-w_{i+2}-1) \\ \frac{1}{2} (w_i+w_{i+1}-w_{i+2}-1)\end{pmatrix}}\ ,
\ee
where the indices are understood mod $3$.
 One then changes variables as in the previous case and expands the integrand around the localizing locus. This leads to the following result for the $y$-integral:
 \be 
 \left| \frac{ \lgamma(k-2)}{\prod_{i=1}^3 \lgamma(2-2j_i)}\prod_{i=1}^3 (\partial_{y_i} X_i)^{3-k-2j_i} \prod_{i<j} (\partial_{y_i} \partial_{y_j} X_{123})^{1-2j_\ell} \prod_{i=1}^3 y_i^{\frac{k w_i}{2}-h_i+j_i-1}\right|^2\ .
 \ee
Here $\ell=6-i-j$ is the third index not equal to $i$ and $j$. This factor should be evaluated at $y_i=a_i$. This can be rewritten entirely in terms of the quantities $a_i$ and $\Pi$ of the covering map thanks to the identities
\begin{subequations}
\begin{align}
\prod_{i=1}^3 \partial_{y_i} X_i&=\prod_{i<j} \partial_{y_i} \partial_{y_j}X_{123}\ , \\
\Pi^6&=\prod_i (w_i a_i)^{-3w_i-1} (\partial_{y_i} X_i)^4\ , \\
(\partial_{y_i} \partial_{y_j} X_{123})^6&=\frac{(w_\ell a_\ell)^2  (\partial_{y_i} X_i)^4(\partial_{y_j} X_j)^4}{(w_i a_i)(w_j a_j)(\partial_{y_\ell} X_\ell)^2}\ ,
\end{align}
\end{subequations}
valid on the locus $y_i=a_i$. We obtain for the residue of the string three-point function
\be 
\frac{C_{\text{S}^2}\nu^{\frac{k}{2}-2}}{2\pi^2(k-2)^2 \lgamma\left(\frac{k-1}{k-2}\right)}\left|\prod_{i=1}^3 a_i^{\frac{k}{4}(w_i-1)-h_i} w_i^{-\frac{k}{4}(w_i+1)+1-j_i} \Pi^{-\frac{k}{2}}\right|^2\ . \label{eq:m=0 string result}
\ee
 \subsection{CFT computation} \label{eq:CFT computation}
 We now show that these results are reproduced by the CFT computation. 
 
 \paragraph{$m=0$.} There is very little work left, since the deformation is actually turned off. We get from \eqref{eq:deformed correlation function} the following expression for the residue on the CFT side
 \be 
 \frac{1}{\pi \sqrt{N} }\prod_{i=1}^3 N(w_i,j_i) w_i^{\frac{1}{2}} \left|\prod_{i=1}^3 a_i^{\frac{k}{4}(w_i-1)-h_i} w_i^{-\frac{k}{4}(w_i+1)} \Pi^{-\frac{k}{2}}\right|^2\ .
 \ee
 Thus provided that
 \be 
 \frac{C_{\text{S}^2}\nu^{\frac{k}{2}-2}}{2\pi^2(k-2)^2 \lgamma\left(\frac{k-1}{k-2}\right)} \prod_{i=1}^3 w_i^{2-2j_i} =\frac{1}{\pi \sqrt{N} }\prod_{i=1}^3 N(w_i,j_i) w_i^{\frac{1}{2}}\ ,
 \ee
 we get agreement with the string answer \eqref{eq:m=0 string result}. This condition determines
 \be 
 N(w_i,j_i)=N_0 w_i^{\frac{3}{2}-2j_i}
 \ee
 and
 \be 
C_{\text{S}^2}=2\pi(k-2)^2 \nu^{2-\frac{k}{2}}\lgamma\left(\frac{k-1}{k-2}\right) N_0^3N^{-\frac{1}{2}} \label{eq:CS2 identification}
 \ee
 in terms of a single undetermined constant $N_0$ (depending only on $k$).
 \paragraph{$m=1$.} For $m=1$, the CFT residue reads
 \begin{multline} 
 \frac{-\sqrt{2}\mu N_0^3 }{\pi \sqrt{N}}\prod_{i=1}^3 w_i^{2-2j_i} \sum_{\begin{subarray}{c} \text{connected} \\ \text{covering surfaces}\end{subarray}} \int \mathrm{d}^2\xi\ \Bigg| 2^{-\frac{3k}{4}}\prod_{i=1}^3 w_i^{-\frac{k(w_i+1)}{4}} a_i^{\frac{k(w_i-1)}{4}-h_i}  \alpha^{\frac{k}{4}-1} \Pi^{-\frac{k}{2}}  \\
\times \ \zeta^{j_1+\frac{k}{2}-2}(\zeta-1)^{j_2+\frac{k}{2}-2}  \Bigg|^2\ .
 \end{multline}
 We used the dictionary \eqref{eq:map of parameters}.
 In this case, there can be several covering surfaces whose number is given by the Hurwitz number $H_{(w_1,w_2,w_3,2)}$.\footnote{For an explicit formula, see \cite{Liu}.} For four-point functions, all these covering spaces are related via analytic continuation in $\xi$ \cite{Liu, Dei:2019iym}. Hence the sum over connected covering surfaces just instructs us to integrate over all possible branches of the implicit functions $z_i(\xi)$, $\zeta(\xi)$, $a_i(\xi)$, $\alpha(\xi)$ and $\Pi(\xi)$. Let us perform the change of variables in the integrand to $a_1$ (that is implicitly determined in terms of $\xi$). Since we will see momentarily that after the change of variables, the integrand is a single-valued function of $a_1$, we exactly cancel the sum over different branches of these multi-valued maps. To perform the change of variables, we need the following identities
 \begin{subequations}
 \begin{align}
X_\emptyset^4&= \frac{\Pi^2 \prod_{j=1}^3 (w_j a_j)^{1+w_j}}{2 \zeta^{2}(\zeta-1)^{2}\alpha} \ , \\
\frac{\partial a_1}{\partial \xi}&=\frac{w_1a_1}{2\zeta^2(\zeta-1)\alpha}\ , \\
\left(\frac{\partial_{y_2} X_{12}}{\partial_{y_2} X_{23}}\right)^2&=\frac{w_1a_1}{\zeta w_3 a_3}\ , \\
\left(\frac{\partial_{y_3} X_{13}}{\partial_{y_3} X_{23}}\right)^2&=\frac{(\zeta-1) w_1a_1}{\zeta w_2 a_2}\ .
 \end{align}
 \end{subequations}
 Thus we obtain for the CFT three-point function
 \begin{align}
 \frac{-\sqrt{2}\mu N_0^3 2^{2-2k}}{\sqrt{N}}\int \frac{\mathrm{d}^2 a_1}{\pi} \left| \prod_{i=1}^3 a_i^{\frac{kw_i}{2}+j_i-h_i-1} X_\emptyset^{-k} \left(\frac{\partial_{y_2} X_{12}}{\partial_{y_2} X_{23}}\right)^{2j_3-2}\left(\frac{\partial_{y_3} X_{13}}{\partial_{y_3} X_{23}}\right)^{2j_2-2}\right|^2\ .
 \end{align}
 This agrees with the string theory result \eqref{eq:m=1 string result} provided that
 \be 
\mu= -\frac{(k-2) 2^{2k-\frac{5}{2}}\nu^{1-\frac{k}{2}}}{\pi}\ . \label{eq:mu identification}
 \ee
 \paragraph{$m=2$.} We observe the following identities:
 \begin{subequations}
 \begin{align}
X_i^4&= \frac{\Pi^2 (\zeta_1-\zeta_2)^2\prod_{j=1}^3 (w_j a_j)^{1+w_j+2\delta_{ij}}}{4 \prod_{\ell=1}^2 \zeta_\ell^{2+2\delta_{i1}}(\zeta_\ell-1)^{2+2\delta_{i2}}\alpha_\ell} \ , \\
\det \frac{\partial (a_1,a_2)}{\partial (\xi_1,\xi_2)}&=\frac{w_1a_1 w_2 a_2(\zeta_1-\zeta_2)}{4\prod_{\ell=1}^2\zeta_\ell^2(\zeta_\ell-1)^2\zeta_2^2\alpha_\ell}\ .
 \end{align}
 \end{subequations}
 We change variables in the integral $(\xi_1,\xi_2) \to (a_1,a_2)$. The integrand becomes again a single-valued function after substitution and thus we compensate the necessary sum over covering maps by this change of variables. However, we also notice that both $a_1(\xi_1,\xi_2)$ and $a_2(\xi_1,\xi_2)$ are by construction symmetric functions (since one can exchange the role of the two marginal operators). Hence we need to accompany the change of variables by a factor of 2 to account for this. This will be also the case at higher orders in perturbation theory and in general this cancels the factor $\frac{1}{m!}$ that is present in \eqref{eq:deformed correlation function}.
 Inserting this in the formula \eqref{eq:deformed correlation function} for the deformed CFT correlator, we obtain at second order for the residue
 \be 
\frac{2^{5-4k}\mu^2}{\sqrt{N}} N_0^3\pi\int \frac{\mathrm{d}^2 a_1 \ \mathrm{d}^2 a_2}{\pi^2} \left| \prod_{i=1}^3 a_i^{\frac{kw_i}{2}+j_i-h_i-1}\prod_{i=1}^2 X_i^{-2j_i} X_3^{2-2j_3} \right|^2\ .
 \ee
 With the identification \eqref{eq:CS2 identification} and \eqref{eq:mu identification}, we get exact agreement with the string result \eqref{eq:m=2 string result}.
\paragraph{$m=3$.} With the help of the identities
\begin{subequations}
\begin{align}
X_{ij}^4&= \frac{\Pi^2 \prod_{1\le \ell<m\le 3}(\zeta_\ell-\zeta_m)^2\prod_{l=1}^3 (w_l a_l)^{1+w_l+2\delta_{l \in \{i,j\}}}}{8 \prod_{\ell=1}^3 \zeta_\ell^{2+2\delta_{1 \in \{i,j\}}}(\zeta_\ell-1)^{2+2\delta_{2 \in \{i,j\}}}\alpha_\ell} \ , \\
\det \frac{\partial (a_1,a_2,a_3)}{\partial (\xi_1,\xi_2,\xi_3)}&=\frac{\prod_{i=1}^3 w_ia_i\prod_{1 \le \ell<m\le 3}(\zeta_i-\zeta_j)}{8\prod_{\ell=1}^3\zeta_\ell^2(\zeta_\ell-1)^2\zeta_2^2\alpha_\ell}\ ,
\end{align}
\end{subequations}
one readily rewrites the CFT expression \eqref{eq:deformed correlation function} as
\be 
-\frac{\mu^3 2^{\frac{15}{2}-6k}N_0^3\pi^2}{\sqrt{N}}\int \prod_{i=1}^3 \frac{\mathrm{d}^2 a_i}{\pi} \ \left|\prod_{i=1}^3 a_i^{\frac{kw_i}{2}+j_i-h_i-1} \prod_{i<j} X_{ij}^{k-2j_i-2j_j}\right|^2\ .
\ee
This matches again with the string result \eqref{eq:D prefactors}, given the identifications \eqref{eq:CS2 identification} and \eqref{eq:mu identification}.

\paragraph{$m=4$.} Finally, we also check the $m=4$ case, where we have more integrals on the CFT side than on the string side. We change variables to
\be 
(\xi_1,\xi_2,\xi_3,\xi_4) \longrightarrow \left(a_1,a_2,a_3,u=\sqrt{\frac{X_1^2 w_3 a_3 \prod_i \zeta_i}{X_3^2 w_1 a_1}}\right)\ .
\ee
We have the following identities:
\begin{subequations}
\begin{align}
X_3(w_3a_3)^{-\frac{1}{2}}&= X_1 (w_1a_1)^{-\frac{1}{2}} \prod_i \zeta_i^{\frac{1}{2}}+ X_2 (w_2a_2)^{-\frac{1}{2}} \prod_i (\zeta_i-1)^{\frac{1}{2}}\ , \\
X_{123}^4&=\frac{\Pi^2 \prod_{1 \le \ell<m \le 4} (\zeta_\ell-\zeta_m)^4 \prod_{i=1}^3 (w_ia_i)^{w_i+3}}{16 \prod_{\ell=1}^4 \zeta_\ell^4(\zeta_\ell-1)^4 \alpha_\ell}\ , \\
\det \frac{\partial(a_1,a_2,a_3,u)}{\partial(\xi_4,\xi_5,\xi_6,\xi_7)} &=\frac{\prod_{1 \le \ell \le m \le 4} (\zeta_\ell-\zeta_m)^{\frac{1}{2}}\prod_{i=1}^2 (w_i a_i)^{\frac{1-w_i}{4}} (w_3 a_3)^{\frac{5-w_3}{4}} X_1 X_2}{8 \prod_{\ell=1}^4 \zeta_\ell(\zeta_\ell-1)\alpha_\ell^{\frac{3}{4}} \Pi^{\frac{1}{2}} X_3}\ .
\end{align}
\end{subequations}
These identities bring the CFT residue into the form
\begin{align}
\frac{\mu^4 2^{10-5k} N_0^3}{\pi \sqrt{N}} \int \prod_{i=1}^3 \mathrm{d}^2 a_i \, \mathrm{d}^2 u \  \left|\prod_{i=1}^3 a_i^{\frac{k w_i}{2}+j_i-h_i-1} \prod_{i=1}^3 X_i^{k-1-2j_i} X_{123}^{1-k} u^{2j_1-k} (1-u)^{2j_2-k}\right|^2 \ .
\end{align}
One can easily perform the $u$-integral and obtains
\be 
\frac{\mu^4 \pi^3 2^{10-5k} N_0^3}{\sqrt{N} \prod_i \lgamma(k-2j_i)}\int \prod_{i=1}^3 \frac{\mathrm{d}^2 a_i}{\pi} \  \left|\prod_{i=1}^3 a_i^{\frac{k w_i}{2}+j_i-h_i-1} \prod_{i=1}^3 X_i^{k-1-2j_i} X_{123}^{1-k} \right|^2 \ .
\ee
This again agrees with the string answer of eq.~\eqref{eq:D prefactors} given the identifications \eqref{eq:CS2 identification} and \eqref{eq:mu identification}.

\subsection{Two-point functions and the reflection coefficient} \label{subsec:two-point function}
We have found perfect agreement for the three-point functions. There is still one undetermined constant in the matching that we will fix by comparing two-point functions.

For two-point functions there are two terms to be matched. Let us start with the simple term, namely the delta-function $b\, \delta(\alpha_1+\alpha_2-Q)$. We can directly match it to the delta-function $\delta(j_1+j_2-1)$ piece in eq.~\eqref{eq:2-point function worldsheet} on the string side. Because of charge conservation, this piece is unchanged in conformal perturbation theory. On the CFT side, vertex operators are canonically normalized. Hence we just have to account for the relative normalization factors $N(w,j)$. Agreement between the string and the CFT result is then the condition
\be 
N(w,j)N(w,1-j)=N_0^2 w=2(k-2) w C_{\mathrm{S}^2}\ .
\ee
This condition together with \eqref{eq:CS2 identification} and \eqref{eq:mu identification} then fixes the three unknown constants and gives \eqref{eq:matching of normalization constants}.

\paragraph{Reflection coefficient.} 
The $\delta(j_1-j_2)$ term of the two-point function is more interesting. To compute it, we assume that the three-point functions have already been matched. This allows us to perform another strong test of the proposed duality that goes beyond conformal perturbation theory. Notice that in the CFT we would have
\be 
\partial_{\mu} \left\langle \cdots \right \rangle= - \int \mathrm{d}^2 \xi \ \left \langle \Phi(\xi) \cdots \right \rangle\ , \label{eq:mu derivative CFT}
\ee
where the dots stand for any number of vertex operators. This follows directly from the appearance of the marginal operator in the action. Let's consider this equality for a two-point function. On the left-hand side, we only pick out the term of the two-point function that contains the reflection coefficient. Thus, we obtain
\be 
\partial_\mu \left\langle \sigma_{w,\alpha_1}(0)\sigma_{w,\alpha_2}(\infty) \right\rangle =\pm i (2\pi)^2 \delta^{(2)}(h_1-h_2) \left \langle\sigma_{w,\alpha_1}(0)\Phi(1)\sigma_{w,\alpha_2}(\infty) \right\rangle\ ,
\ee
where we used conformal invariance to fix the $\xi$-dependence and perform the integral.
The right-hand side is a special case of the three-point function that we already matched with the worldsheet. Thus we arrive at the prediction that that the special three-point function with $j_1=j_2=j$, $h_1=h_2=h$, $w_1=w_2=w$ and $j_3=3-k$, $h_3=1$, $w_3=2$ should reduce to the derivative of the two-point function. Using the explicit representation of the three-point function \eqref{eq:3-point function worldsheet}, one calculates for this special three-point function (see Appendix~\ref{app:reduction two-point function} for the calculation)
\begin{multline} 
\left \langle V_{j,h}^{w}(0;0)V_{j,h}^{w}(1;1)V_{j=3-k,h=1}^{w=2}(\infty;\infty) \right \rangle=\frac{(2j-1)^2}{8\pi^2\nu^{2j+2-k}\lgamma\left(\frac{k-1}{k-2}\right) (k-2)^4 \lgamma\left(\frac{2j-1}{k-2}\right)} \\
\times \frac{\lgamma\left(h+j-\frac{kw}{2}\right)}{\lgamma\left(h-j-\frac{kw}{2}+1\right)\lgamma(2j)}\ . \label{eq:special worldsheet 3-point function}
\end{multline}
Given the matching between three-point functions, this implies now that
\begin{multline} 
\partial_\mu \left \langle\sigma_{w,\alpha_1}(0)\sigma_{w,\alpha_2}(\infty)\right\rangle=\frac{C_{\mathrm{S}^2}}{ N(w,j)^2N(w,3-k)} \frac{2\pi^2(k-2) w}{(2j-1)} \delta(j_1-j_2)\\
\times \left \langle V_{j,h}^{w}(0;0)V_{j,h}^{w}(1;1)V_{j=3-k,h=1}^{w=2}(\infty;\infty) \right \rangle\ .
\end{multline}
Using the dictionary \eqref{eq:matching of normalization constants}, we can integrate and find
\be 
\left \langle\sigma_{w,\alpha_1}(0)\sigma_{w,\alpha_2}(\infty)\right\rangle=w^{4j-2} R(j,h,\bar{h})\ ,
\ee
where we wrote the result in terms of $\nu$. This agrees with the direct worldsheet computation
\be 
\frac{C_{\text{S}^2}}{N(w,j)^2} 2w(k-2) R(j,h,\bar{h})\ .
\ee
This computation shows that at least on the level of two- and three-point function, the exponential potential that we have assumed is also realized beyond conformal perturbation theory.

\subsection{Fusion rules}
After having computed three-point functions on both sides of the proposed correspondence, let us take a step back and focus on the rougher structure of the answer by looking at the fusion rules. 

Let us first recall that the fusion rules of the undeformed symmetric orbifold side follow directly from the Riemann-Hurwitz formula \eqref{eq:Riemann-Hurwitz formula} that gives the necessary condition for the existence of a covering map. It implies
\be 
\sum_i w_i \ge 1+2 \max_i w_i\ , \qquad \sum_i w_i \in 2\mathbb{Z}+1\ .
\ee
In conformal perturbation theory, one is deforming by a twist-2 operator, which can modify the fusion rules. A priori, one could perhaps expect that the three-point function for any choice of $(w_1,w_2,w_3)$ can be non-zero. This is not the case, but the generic selection rules become weaker:
\be 
\sum_i w_i\ge -1+2 \max_i w_i\ .  \label{eq:deformed fusion rules}
\ee 
Even though the parity constraint is gone, the answer still depends significantly on it as one can see in the explicit formula \eqref{eq:3-point function worldsheet}.

The existence of any selection rule is perhaps surprising from the CFT point of view, given that any
\be 
\left\langle \prod_{i=1}^3 \sigma_{w_i,\alpha_i}(x_i) \prod_{\ell=1}^m \Phi(\xi_\ell) \right \rangle
\ee
are non-zero for high enough $m$. One only gets a vanishing result after integrating over $\xi_\ell$. One can see this explicitly in the computation above, although we have not emphasized it.
For example, let us reconsider third order conformal perturbation theory and let us put $w_3=w_1+w_2+2$. In this case, the integrand is certainly non-vanishing. After the same change of variables that we performed above, we can write the integral simply as
\be 
\int \prod_{i=1}^3 \mathrm{d}^2 a_i \ \left| a_1^{\frac{kw_1}{2}-h_1+j_1-1}a_2^{\frac{kw_2}{2}-h_2+j_2-1}a_3^{\frac{kw_3}{2}-h_3-j_3-1}\right|^2\ ,
\ee
which is in accordance with the worldsheet answer \eqref{eq:3-point function worldsheet}, given that $X_{12}=1$, $X_{13}=y_3$, $X_{23}=y_3$ in this case ($X_\emptyset$ does not appear because of the constraint $j_1+j_2+j_3=k$). 
After integration, this gives delta-functions that impose the additional conditions beyond $\sum_i j_i=k$
\be 
h_1=j_1+\frac{kw_1}{2}\ , \qquad h_2=j_2+\frac{kw_2}{2}\ , \qquad h_3=-j_3+\frac{kw_3}{2}\ .
\ee
Hence generically these three-point functions vanish. If the bound \eqref{eq:deformed fusion rules} is violated even more, the three-point functions always have to vanish.

 \section{Poles in correlators and short strings} \label{sec:poles and short strings}
 Let us recap what we showed so far. We showed that there is a CFT whose two- and three-point functions agree with the prediction from string theory. So far, we only have talked about correlation functions of continuous series representations on the worldsheet which get mapped to delta-function normalizable operators in the dual CFT.
 
This is however not the full story, since there are also discrete series representations on the worldsheet, and they should get mapped to normalizable operator in the proposed boundary CFT. These normalizable operators should be thought of as `bound states' that exist thanks to the introduction of the exponential wall in the symmetric orbifold. One can discover them by analyzing poles in the correlation functions.

 Following \cite{Aharony:2004xn}, let us first review what kind of poles one should expect in correlation functions. There are two different kind of poles that have different physical origin.
 
 \subsection{LSZ poles}
 The first type of poles are the so-called LSZ poles. One should think of them as as the analogue of the familiar LSZ poles in QFT, where an external particle goes on-shell. In the present context, every vertex operator has two labels: $j$ (related to the momentum via \eqref{eq:map of parameters}) and $h$. While these two parameters are in principle related by the mass-shell condition \eqref{eq:mass shell condition solution h}, this relation can always be modified by dressing the vertex operator with some operator in the internal CFT. Hence for the present purposes, we may assume $j$ and $h$ to be independent. 
 
Because of the interpretation of LSZ poles, they appear as poles when tuning $h$ and $j$ of a single vertex operator. In fact, it is easy to see from the structure of two- and three-point functions \eqref{eq:2-point function worldsheet} and \eqref{eq:3-point function worldsheet} that LSZ poles appear whenever
\be 
h-\frac{kw}{2}-j \in \mathbb{Z}_{\ge 0} \quad \text{or}\quad h-\frac{kw}{2}+j \in \mathbb{Z}_{\le 0}\ , \label{eq:LSZ poles}
\ee
corresponding to a divergence from the $y_i \sim 0$ or the $y_i \sim \infty$ region in the integral in eq.~\eqref{eq:3-point function worldsheet}. It was shown in \cite{Dei:2021yom}, that the representation of correlators in terms of $y$-integral is also possible for four-point and higher-point functions. Thus the condition for the appearance of LSZ poles is the same for higher-point functions, as suggested by their interpretation of a particle going on-shell.

This thus suggests the existence of further operators in the CFT that satisfy the LSZ pole condition \eqref{eq:LSZ poles}. These operators will be dual to short string states on the worldsheet. Importantly, not all LSZ poles that appear in correlation functions correspond to normalizable operators. Below we will identify those poles that are associated with normalizable operators.

Let us also remark that the LSZ pole condition \eqref{eq:LSZ poles} is somewhat unusual from a standard CFT point of view. After inserting the mass-shell condition \eqref{eq:mass shell condition solution h}, it reads
\be 
-\frac{j(j-1)}{w(k-2)}-\frac{k}{4}w+\frac{\Delta_X-1}{w}-j \in \mathbb{Z}_{\ge 0} \ ,
\ee
and similarly for the second possibility. Here we included a contribution from the internal vertex operator $\Delta_X$. This is a quadratic condition on the spin $j$ and hence leads to a highly non-trivial discrete spectrum. While its appearance from the worldsheet is well-known, we emphasize that it appears here directly from the CFT without need to refer to the string worldsheet.  
\subsection{Bulk poles} \label{subsec:bulk poles}
The second type of poles are bulk poles. These are actually the poles that we used in Section~\ref{sec:matching} and whose residues can be computed in conformal perturbation theory. However we only used a subset of them. 
Let us again explain their physical origin and be more complete about the full set of poles. In the CFT, we saw that the path integral is convergent whenever 
\be 
\Re \left(Q-\sum_i \alpha_i \right)>0\ .
\ee
If $Q=\sum_i \alpha_i$, we observe the first bulk pole. For this value of the spins, $\phi$ has a zero mode, which leads to the divergence. Hence the bulk pole is intimately related to the infinite volume of $\mathrm{AdS}_3$.

We also saw that this inequality can be violated by any number of marginal operators, i.e.\ we also have bulk poles for
\be 
\sum_i \alpha_i=Q +\frac{m}{2b}
\ee
with $m \in \mathbb{Z}_{\ge 0}$. Also, the parity constraint in the symmetric orbifold requires that $m=\sum_i(w_i-1) \bmod 2$. For a three-point function, this means
\be 
j_1+j_2+j_3=2+\frac{1}{2}(k-2)(m-1)\ .
\ee
But these cannot be the only poles. We can see this as follows. The vertex operators in our normalization take the form $\mathrm{e}^{-2 \alpha_i \phi}(x)$. 
Hence the zero mode path integral for a correlation function at $m$-th order in conformal perturbation theory becomes
\be 
\int \mathrm{d}\phi \ \mathrm{e}^{\left(2Q-\frac{m}{b}-2\sum_i \alpha_i\right) \phi} \left(1+ \mathcal{O}(\mathrm{e}^{-2b \phi}) \right) \ .
\ee
There is a whole tower of subleading corrections. They lead to the resonances
\be 
\sum_i \alpha_i=Q+\frac{m}{2b}+bn
\ee
for $m$ as above and $n \in \mathbb{Z}_{\ge 0}$. Finally, the theory enjoys a reflection symmetry that relates the vertex operator with momentum $\alpha$ to its reflected counter part with $Q-\alpha$. For delta-function normalizable operators, these two operators are identified, whereas for the discrete spectrum at most one of these operators is normalizable. 

Overall, we explained that the boundary CFT three-point functions have the following bulk poles:
\be 
j_1+j_2+j_3-1 =\frac{1}{2}(k-2)(m-1)+n\ , \label{eq:bulk poles}
\ee
where $m\in \mathbb{Z}_{\ge 0}$, $n \in \mathbb{Z}_{\ge 1}$ and $m=\sum_i(w_i-1) \bmod 2$.\footnote{We shifted the definition of $n$ by one to conform with standard string theory conventions.} Additionally, there are poles for any $j_i$ replaced by $1-j_i$. In Appendix~\ref{app:string bulk poles}, we show that the string theory three-point function \eqref{eq:3-point function worldsheet} has the same set of poles, thus giving further evidence for the matching.

There is a small subtlety. Three-point functions with $2\max_i(w_i+1)=\sum_i(w_i+1)$ are vanishing in the symmetric orbifold, but non-vanishing in the deformed symmetric orbifold. One can easily see that they first receive contributions at second order in conformal perturbation theory. Thus, we additionally need to require
\be 
m \ge 1+2 \max_i w_i-\sum_i w_i
\ee
in \eqref{eq:bulk poles} to get the correct set of bulk poles. In Appendix~\ref{app:string bulk poles}, we show that this is also the case in the explicit worldsheet formula \eqref{eq:3-point function worldsheet}.
 
\subsection{Identification of discrete states} \label{subsec:discrete spectrum}
We have seen that the LSZ poles give rise to the discrete representations. However, we have not found out yet which operators are normalizable and which are non-normalizable from the CFT perspective. 

To start with, we should recall that delta-function normalizable operators are a superposition of the incoming and outgoing waves. For discrete operators, either only the incoming or outgoing wave is normalizable. In our convention, vertex operators with $\alpha>\frac{Q}{2}$ (i.e.\ $j>\frac{1}{2}$) are normalizable, as can be seen from the action \eqref{eq:deformed action}. 

However, this is not the whole story. To continue, we argue directly from the CFT that there is an identification of operators
\be 
w^{\frac{3}{2}-2j} (w+1)^{-\frac{3}{2}+k-2j}\mathcal{N}(j) \sigma_{w,j,h=j+\frac{kw}{2}}=\sigma_{w+1,\frac{k}{2}-j,h=j+\frac{kw}{2}} \label{eq:identification of discrete operators}
\ee
in the CFT. This identification is the analogue to the identification of representations on the worldsheet \eqref{eq:D+D- identification}. The identification only makes sense for $w\ge 1$, since a twist-0 operator has no meaning in the CFT. However, this identification will allow us to identify certain $w=1$ twist operators with $w=0$ short string states.
To show this, we insert these operators in three-point functions. It is a simple computation to show that
\begin{multline} 
\Res_{h_3=j_3+\frac{kw_3}{2}} \langle V_{j_1,h_1}^{w_1}(0;0) V_{j_2,h_2}^{w_3}(1;1)V_{j_3,h_3}^{w_3}(\infty;\infty)\rangle\\
=\mathcal{N}(j)\Res_{h_3=j_3+\frac{kw_3}{2}}  \langle V_{j_1,h_1}^{w_1}(0;0) V_{j_2,h_2}^{w_3}(1;1)V_{\frac{k}{2}-j_3,h_3}^{w_3+1}(\infty;\infty)\rangle
\end{multline}
The additional prefactors in \eqref{eq:identification of discrete operators} come from the different normalization of the operators in the CFT as specified by $N(w,j)$. Thus the difference of the LHS and the RHS of \eqref{eq:identification of discrete operators} vanishes in any three-point function (at least for the subset we have computed). This suggests the validity of the equality in the CFT. 

Now we see that there is a stronger requirement for normalizability, namely the operator should be normalizable in both ways of writing. Hence both $j>\frac{1}{2}$ and $\frac{k}{2}-j > \frac{1}{2}$, which leads to the necessary condition
\be 
\text{Normalizability:}\qquad \frac{1}{2}<j<\frac{k-1}{2}\ .
\ee
We do not have a CFT argument that implies that this condition is also sufficient.
At this point, we have hence reproduced the entire worldsheet spectrum from the dual CFT. Owing to the identification \eqref{eq:identification of discrete operators}, there are also discrete operators in the spectrum that have formally $w=0$. But from the CFT point of view, it is more natural to consider them as operators in the $w=1$ sector.

\subsection{Identity operators and the shadow transform}
There are two several special operators in the discrete spectrum that deserve special attention.

\paragraph{Identity operator.} From the worldsheet analysis, it is well-known that there is a unique operator with $h=0$ (for $k>3$). It is mapped to the dual CFT operator
\be 
\mathcal{I}=\sigma_{w=1,j=\frac{k}{2}-1}\ ,
\ee
which indeed has $h=0$. In terms of the momenta, it satisfies $\alpha=Q$. It is hence to be identified with the vacuum of the spacetime CFT. Notice that it satisfies also $h-\frac{k}{2}+j=-1\in \mathbb{Z}_{\le 0}$ and hence appears as an LSZ pole in correlation functions.
This operator is normalizable for $k>3$ and $k<3$. Thus the theory with $k<3$ does not have a normalizable vacuum state \cite{Giveon:2005mi}.\footnote{There seems to be yet another choice for the vacuum, namely the reflected version with $\alpha=0$ (corresponding to $j=2-\frac{k}{2}$) and $w=1$. This operator becomes normalizable for $k<3$. However, it does not sit on an LSZ pole \eqref{eq:LSZ poles}, since $h-\frac{k}{2}-j=-2<0$ and thus does not appear as an operator in the discrete spectrum.}
\paragraph{Marginal operator.} There is a second important operator in our analysis, which is given by
\be 
\Phi=\sigma_{w=2,j=3-k}\ ,
\ee
which is the marginal operator ($\alpha=-\frac{1}{2b}$ in the language of momenta). It has conformal weight 1 and is non-normalizable for $k>\frac{5}{2}$.\footnote{The inequality $k>\frac{5}{2}$ can be violated in the bosonic string, but not in the superstring. We take this as indication that this bound does not have an important physical meaning.} This operator should be non-normalizable because it adds an exponential wall to the theory. We should also mention that it is \emph{not} a local operator of the theory, since it does not lead to an LSZ pole. Indeed, we have
\be 
h-\frac{kw}{2}-j=-2<0\ ,
\ee
and hence the condition \eqref{eq:LSZ poles} is not satisfied. We are thus deforming by an operator that is not in the spectrum of the theory under consideration! 

\paragraph{Shadow transform.} It is perhaps puzzling that there are two operators that behave very similarly. It is well-known from the literature that $\mathcal{I}$ is to be identified with the zero-mode of the dilaton \cite{Giveon:1998ns, Kim:2015gak}. Indeed, inserting it into a worldsheet correlation function has the effect of a $\nu$-derivative on the worldsheet. We saw by explicit computation in Section~\ref{subsec:two-point function} that the integrated $\Phi$ operator has exactly the same effect on the worldsheet. Thus, up to normalization, we should identify
\be 
\mathcal{I}(x)\sim \int \mathrm{d}^2 \xi\ \Phi(\xi)\ . \label{eq:shadow transform identification}
\ee
This makes sense, since the identity operator is independent of $x$. This is a special case of the shadow transform and we have hence argued that $\mathcal{I}$ and $\Phi$ are shadow transforms. In particular, $\mathcal{I}$ is a local operator and $\Phi$ is its non-local shadow.

\section{Grand canonical ensemble and large \texorpdfstring{$\boldsymbol{N}$}{N} expansion} \label{sec:grand canonical ensemble}
There is one confusing aspect of our proposal that we now address. Due to the identification \eqref{eq:shadow transform identification}, 
it seems like we have just added the identity operator to our theory and hence not changed the theory at all. The resolution to this puzzle is known in the literature \cite{Kim:2015gak} and we now explain its interpretation in the context of this duality.

\subsection{Grand canonical ensemble}
The operator $\mathcal{I}(x)$ is not the identity operator in the CFT sense. In fact, it was found on the worldsheet
that inserting $\mathcal{I}(x)$ in a correlation function is a non-trivial operation \cite{Giveon:1998ns, Kim:2015gak}. Since $\mathcal{I}(x)$ is the dilaton zero-mode, we have up to normalizations
\be 
\langle \mathcal{I}(x) \cdots \rangle \sim \partial_\mu  \langle \cdots \rangle\ ,
\ee
where the dots stand for any additional vertex operators.
While the $\mu$-dependence of a correlator is fixed by KPZ scaling \eqref{eq:KPZ scaling}, the exponent depends on the momenta of the other vertex operators. So while the LHS is proportional to the RHS, the proportionality factor depends on the other vertex operators and hence $\mathcal{I}(x)$ cannot be the identity operator in the CFT sense. 

From the CFT point of view $\int \mathrm{d}^2 \xi \ \Phi(\xi) \sim \mathcal{I}$ is the deforming operator and hence it is clear that this should happen. In fact, this is eq.~\eqref{eq:mu derivative CFT}. Adding the marginal operator can be simply viewed as performing a Legendre transform of the CFT, where instead of keeping $\langle \mathcal{I} \rangle$ (and hence the central charge) fixed, we fix a corresponding chemical potential $\mu$. Indeed, on the level of the path integral the deformation of the symmetric orbifold is
\be 
\int \mathscr{D}[\text{fields}] \ \mathrm{e}^{-S} \longrightarrow \int \mathscr{D}[\text{fields}] \ \mathrm{e}^{-S+ \mu \mathcal{I}}\ .
\ee
In the saddle-point approximation to the path integral, this is exactly the definition of a Legendre transform.

After this operation, the dual CFT is not quite a CFT anymore, because we have traded the parameter $N$ that specified the central charge of the undeformed CFT for $\mu$. We saw that $N$ could be fully absorbed into the definition of $\mu$ in eq.~\eqref{eq:mu identification}. $N$ is no longer an independent parameter in the deformed CFT, at least at leading order in $N^{-1}$. $\mu$ in turn gets identified with the only free parameter on the string side -- the string coupling, see eq.~\eqref{eq:identification string coupling} below for the precise statement. The CFT should then be viewed as a grand canonical ensemble of CFTs.

\subsection{Large \texorpdfstring{$N$}{N} expansion} \label{subsec:large N}
Let us now also address a related matter. For fixed $N$, the symmetric orbifold can only compute correlators as long as the degree of the relevant covering map is less than $N$. In view of the Riemann-Hurwitz formula \eqref{eq:Riemann-Hurwitz formula}, this means that at $m$-th order in conformal perturbation theory and at leading order in $\frac{1}{N}$, we have the constraint 
\be 
d=1+\frac{1}{2}\sum_i(w_i-1)+\frac{m}{2}\le N\ ,
\ee
and so $m$ is bounded from above.
Thus, one actually finds that only finitely many orders in conformal perturbation theory can contribute for any finite $N$ and fixed genus. This leads to a CFT with radically different properties than what we want, since it is no longer possible to analytically continue the conformal perturbation series to all orders and get a meromorphic answer in the spins for the correlation functions.
For example, there would be a selection rule on correlators that ensures that the $\mu$ exponent is always a positive integer that is bounded from above by the maximal order of the perturbation series. 

This is obviously not the theory that we want. Hence we need to reconsider how exactly we define the dual CFT. Since only an infinite conformal perturbation theory series can possibly lead to an answer for the three-point functions that is meromorphic in the spins, we only know how to define our dual CFT perturbatively in $N^{-1}$ in the strict asymptotic regime $N \to \infty$. In lieu of a better definition of the CFT, we take $N \to \infty$ and hence the perturbation series does not truncate and we may also expand any possible combinatorial prefactor in $\frac{1}{N}$.

\paragraph{Vacuum diagrams.} Let us reconsider the leading large $N$ contribution to a correlator on the sphere. The story is slightly subtler than what we described in Section~\ref{subsec:symmetric product orbifold}, because momentum conservation cannot be used beyond perturbation theory. In particular, there can be any number of disconnected components in the covering spaces and while $\langle \prod_{\ell=1}^m \mathrm{e}^{-\frac{\phi}{b}}\rangle$ vanishes on any Riemann surface for every $m$ in the undeformed theory, this does not mean that it would vanish in the full interacting theory. In fact, KPZ scaling implies that the genus $g$ partition function should scale like $\mu^{-2b Q(1-g)}$, which is generically never a positive integer and hence invisible in perturbation theory. These vacuum diagrams should map to the vacuum diagrams in string theory. They are present for any correlator.

\paragraph{Higher genus corrections.} A connected, normalized correlator receives contributions from higher genus covering maps. On these higher genus covering maps, the KPZ scaling is different and we have instead of eq.~\eqref{eq:KPZ scaling}
\be 
\mu^{2b(\sum_i \alpha_i- Q(1-g))} N^{1-g-\frac{n}{2}}\label{eq:mu dependence higher genus}
\ee
as the $\mu$ and $N$-dependence. This dependence on $\mu$ and $N$ is mirrored on the worldsheet, which predicts\footnote{The normalization of the path integral at higher genus has to be proportional to $C_{\mathrm{S}^2}^{1-g}$, since this is the only ambiguity in the normalization of the string path integral.}
\be 
\frac{C_{\mathrm{S}^2}^{1-g}\nu^{1-g-\sum_i j_i}}{\prod_i N(j_i)}  \sim \nu^{(1-g)(k-3)-\sum_i (j_i+\frac{k}{2}-2)} N^{1-g-\frac{n}{2}}\ , \label{eq:identification string coupling}
\ee
which via the identification of $\mu$ with $\nu$ in \eqref{eq:matching of normalization constants} translates to \eqref{eq:mu dependence higher genus}.

This shows explicitly that $N$ is no longer an independent parameter of the theory, since it can be absorbed into the normalization of the partition function and the vertex operators, as happens on the worldsheet.

We also note that the string coupling is
\be 
g_\text{string}^2\sim \frac{1}{\nu\,  C_{\text{S}^2}}\sim \nu^{3-k} N^{-1} \sim \mu^{2b Q} N^{-1}=\mu^{\frac{2(k-3)}{k-2}} N^{-1}\ .
\ee
Thus we again see the qualitative difference between $k<3$ and $k>3$. $k<3$ backgrounds are weakly coupled for small $\nu$ (large $\mu$), whereas $k>3$ backgrounds are weakly coupled for large $\nu$ (small $\mu$). Of course since also $N$ is large is suffices to just keep $\mu$ finite to ensure small string coupling independently of whether $k<3$ or $k>3$.

We also mention that the different $\mu$-dependence at higher genus changes the location of the bulk poles. The same logic as in Section~\ref{subsec:bulk poles} predicts the following set of bulk poles:
\be 
j_1+j_2+j_3-1=\frac{1}{2}(k-2)(m-1-2g)+g+n
\ee
where $m \in \mathbb{Z}_{\ge 0}$, $n \in \mathbb{Z}_{\ge 1}$ and $m=\sum_i (w_i-1) \bmod 2$.

\section{Superstrings} \label{sec:superstrings}
For technical simplicity we have considered the bosonic string. However, as is of course well-known the bosonic string suffers from the presence of the tachyon and thus loop amplitudes are not well-defined. Hence to get a well-defined duality, we should consider the superstring. We will consider here type the IIB string on $\mathrm{AdS}_3 \times \mathrm{S}^3 \times X$, where $X=\mathbb{T}^4$ or $\mathrm{K3}$. We expect that one can formulate similar conjectures for other supersymmetric backgrounds as well. We change our conventions such that the supersymmetric level of the background is $k$, which leads to various shifts $k \to k+2$ in the bosonic formulas. For the superstring, $k$ is an integer, because of the presence of the compact $\mathrm{S}^3$ factor.

\subsection{Undeformed symmetric orbifold}
The natural candidate orbifold that describes the asymptotic region of $\mathrm{AdS}_3$ is
\be 
\text{Sym}^N\left(\mathbb{R}_Q \times \mathfrak{su}(2)_{k-2} \times \text{four free fermions} \times X\right)\ . \label{eq:N=4 symmetric orbifold}
\ee
Here,
\be 
Q=b^{-1}-b=\frac{k-1}{\sqrt{k}}\ ,\qquad b=\frac{1}{\sqrt{k}}\ .
\ee
The map from spin on the worldsheet to momenta in the linear dilaton factor is
\be 
\alpha=\frac{j+\frac{k}{2}-1}{\sqrt{k}}\ .
\ee
The seed theory of the symmetric orbifold should be thought of as an $\mathcal{N}=4$ linear dilaton theory combined with the internal $X$. This part of the duality was analyzed in detail in \cite{Eberhardt:2019qcl}. The seed theory enjoys $\mathcal{N}=4$ supersymmetry and has the right spectrum of delta-function normalizable operators. 
\subsection{Marginal deformation}
The natural analogue of the bosonic marginal operator is a supersymmetric twist-2 operator. In any $\mathcal{N}=4$ theory, marginal operators are obtained as descendants of BPS operators with $h=\bar{h}=\frac{1}{2}$. In our case, we can write the marginal operator as follows:
\be 
 \Phi \equiv G_{-\frac{1}{2}}^{ \alpha A} \bar{G}_{-\frac{1}{2}}^{\beta B} \Psi_{\alpha\beta,AB}\ . \label{eq:N=4 marginal operator}
\ee
Here, $\alpha$ and $\beta$ are R-symmetry spinor indices and $A$ and $B$ are spinor indices of the outer automorphism group of $\mathcal{N}=4$ supersymmetry. $\Psi_{\alpha\beta,AB}$ is the BPS operator. Contrary to local unitary CFTs, $\Psi_{\alpha\beta,AB}$ transforms as a doublet under the two outer automorphism groups so that the marginal operator is a singlet. This is not in contradiction with standard lore because $\Psi_{\alpha\beta,AB}$ is not a local operator.

We choose $\Psi_{\alpha\beta,AB}$ to be in the twist-2 sector and dress it again by a vertex operator in the linear dilaton direction. We also remember that the fermions are Ramond-moded in the twist-2 sector. The Ramond ground state energy is $\frac{8}{16}=\frac{1}{2}$. Hence the conformal weight of such an operator is
\be 
h=\frac{3k}{8}+\frac{1}{2}\alpha(Q-\alpha)+\frac{8}{2 \times 16}\ ,
\ee
which for $\alpha=-\frac{1}{2b}$ (i.e.\ $j=1-k$) has $h=\frac{1}{2}$. As always the Ramond ground state is degenerate. Focusing on left-movers, the ground state transforms under the $\mathfrak{su}(2)_\text{R} \times \mathfrak{su}(2)_\text{outer}$ as $(\mathbf{3},\mathbf{1}) \oplus (\mathbf{2},\mathbf{2}) \oplus (\mathbf{1},\mathbf{1})$. Additionally, there might be more Ramond ground states that depend on the specific $X$. For example $h^{1,1}(\mathrm{K3})=20$, which leads to more ground states.\footnote{More precisely, the ground states that we have listed come from $h^{0,0}(X)=1$ and $h^{2,0}(X)=1$, which are always present because $X$ is hyperk\"ahler.}  We observe that there is one bi-spinor representation $(\mathbf{2},\mathbf{2})$ that is always present, which is our candidate for the BPS state.

We thus see that there is a unique candidate marginal operator that we can add to the theory. Aside from some additional dressings with free fields, this operator is the same as in the bosonic theory. In particular it still has momentum $\alpha=-\frac{1}{2b}$ and lives in the $w=2$ sector. In terms of the worldsheet spin, the operator has $j=1-k$, which is the same as in the bosonic case after a replacement $k \to k+2$.
\subsection{Matching with the worldsheet computation}
We have not attempted to perform a detailed matching of the correlators of this theory to the worldsheet correlators of superstrings on $\mathrm{AdS}_3 \times \mathrm{S}^3 \times X$, since this is complicated by the presence of the additional free fields. Instead, we list here the qualitative matching that makes it plausible that the matching should extend to the superstring case.
\begin{enumerate}
\item The pole structure in the three-point functions is correct. This follows immediately from the fact that we matched both the LSZ-poles and the bulk poles in the bosonic string and the deforming operator here has the same relevant quantum numbers. The dressing by additional free fields does not change this, since the free fields will never give rise to poles in the correlators.
\item As a consequence the discrete spectrum of the proposed CFT matches the worldsheet spectrum.
\item We also expect additional dressings from the worldsheet perspective (in the RNS formalism). $\int \mathrm{d}^2 \xi \ \Phi(\xi)$ is supposed to correspond to the dilaton zero-mode. On the worldsheet, every operator has a picture number. On the sphere, there should be two operators with picture number $-1$ and the remaining operators with picture number $0$. If we insert the worldsheet avatar of $\Phi$ in a correlation function then we should insert the picture number $0$ version of it to keep the picture number balanced. But the picture number $0$ version is obtained by acting with $G_{-\frac{1}{2}}$ on the picture number $-1$ version.\footnote{There are also other terms, but they do not matter for correlation functions on the sphere.} We view the additional action of the supercharge as introducing a similar dressing of the operator as the dressing $\Psi_{\alpha\beta,AB}$ by the $\mathcal{N}=4$ supercharges in \eqref{eq:N=4 marginal operator}.
\end{enumerate}
\subsection{The tensionless limit}
A special case of superstrings on $\mathrm{AdS}_3 \times \mathrm{S}^3 \times \mathbb{T}^4$ occurs for the minimal level $k=1$. In this case, the dual CFT is conjectured to be the usual symmetric orbifold of $\mathbb{T}^4$, without the need to turn on a marginal deformation \cite{Eberhardt:2018ouy}. Here we explain how this statement is compatible with what we discussed in this paper. We observe that in this case $Q=0$ and simultaneously, the level of the $\mathfrak{su}(2)$ theory in \eqref{eq:N=4 symmetric orbifold} becomes negative and thus the proposal no longer leads to a unitary CFT. However, there is a natural limit of the $\mathcal{N}=4$ dilaton theory for $k \to 1$. Unitarity implies that every operator in the $\mathcal{N}=4$ theory 
\be 
\mathbb{R} \times \mathfrak{su}(2)_{-1} \times \text{4 free fermions}
\ee
is null and hence this theory is in fact trivial. This can also be seen more explicitly by using the free-field realization of $\mathbb{R} \times \mathfrak{su}(2)_{-1}$ in terms of two $\beta\gamma$-ghosts, which in turn cancel the 4 free fermions. Thus, the starting point \eqref{eq:N=4 symmetric orbifold} becomes simply the symmetric orbifold of $\mathbb{T}^4$. Since it does not possess a non-compact direction, there is no Liouville type deformation that one can turn on and this is the exact dual CFT. Correspondingly, two- and three-point functions are far simpler in this theory than what we discussed here, but they are still non-trivial to extract from the worldsheet CFT. 

Since no deformation is needed to define the dual CFT, this proposal is non-perturbatively well-defined in the tensionless limit and one can use it to investigate non-perturbative phenomena, as was done in \cite{Eberhardt:2020bgq, Eberhardt:2021jvj}. We do not understand currently whether this is possible for the proposal made in this paper.

\section{Discussion} \label{sec:discussion}
Let us finally discuss possible issues of our proposal and promising future directions.

\paragraph{Non-perturbative definition of the CFT.} Our construction of the dual CFT was ad hoc. We defined it in perturbation theory in the deforming parameter $\mu$. However, since the deforming operator is non-normalizable, this conformal perturbation theory series is not well-behaved. In particular to get the desired three-point function, one has to analytically continue the perturbation series. This requires in particular that one takes $N \to \infty$, since otherwise there would be a cutoff on the order of perturbation theory that is allowed to contribute. It would be highly desirable to give a more first-principle definition of the CFT and calculate observables using other methods than conformal perturbation theory. In particular, it might be possible to give a semiclassical evaluation of correlation functions similar to Liouville theory \cite{Seiberg:1990eb, Harlow:2011ny, Honda:2013pba} or it could be possible to give a more axiomatic definition of the CFT and extract the two- and three-point functions from bootstrap constraints as was done in \cite{Teschner:1995yf, Teschner:2001rv} for Liouville theory. Such a computation could perhaps also be done at finite $N$, thus possibly leading to a non-perturbative understanding of the duality.

\paragraph{Uniqueness of the two- and three-point functions.} A more direct concern is whether the worldsheet three-point function \eqref{eq:3-point function worldsheet} is the unique answer that is compatible with the residues that are computed from conformal perturbation theory. In other words, one may wonder whether we have uniquely defined the theory. In Liouville theory, the answer to this question is `yes', but after some additional inputs. For example, Liouville theory is invariant under the replacement $b \to b^{-1}$ and the resulting three-point function has to exhibit this symmetry. No such additional symmetry is known to exist for the theory at hand and the question becomes harder to answer. 

\paragraph{Discrete normalizable spectrum.} We gave arguments that determined the discrete normalizable spectrum of the CFT in Section~\ref{subsec:discrete spectrum}. These arguments were guided by the worldsheet theory and only gave necessary conditions for the normalizability of operators. It would be desirable to gain a better understanding of the discrete normalizable states and strengthen our arguments.

\paragraph{$k<3$ vs.\ $k>3$.} In \cite{Balthazar:2021xeh, Martinec:2021vpk} a similar proposal to ours was made in a more restrictive setting for backgrounds with $k<3$ ($k<1$ in the supersymmetric language). The authors restricted themselves to these backgrounds, because the sign of $Q$ is negative for this region. This property is desirable since it leads to a weak coupling region far away from the exponential wall and the theory can be defined more directly in terms of a converging path integral. Another important difference between the $k<3$ and $k>3$ backgrounds is that in the latter the identity operator $\mathcal{I}$ is part of the normalizable spectrum, whereas it is not in the former.
We have seen directly in our analysis that there is no problem in leaving the region $k<3$. The dual CFT that we have constructed is perfectly well-defined for both regions (at least perturbatively in $N^{-1}$) and leads to the correct correlation functions, thus essentially proving the correspondence. 

\paragraph{Euclidean vs.\ Lorentzian signature.} We actually considered the worldsheet theory that describes stings in Euclidean $\mathrm{AdS}_3$. Correspondingly, we took the dual CFT to be Euclidean. The correspondence seems far more natural in this setting, since it relies heavily on orbifold technology particular to Euclidean signature, such as the existence of branched covering maps. We currently do not understand very well how to interpret our results in Lorentzian signature.

\paragraph{Four-point function.} Beyond the matching of two- and three-point functions, it would be interesting to also compare the four-point functions. On the worldsheet, there is a proposed closed form of these correlation functions, which one can in principle similarly check order by order in the deformation parameter $\mu$. We expect that for this the $H_3^+$/Liouville correspondence will be very useful \cite{Ribault:2005wp, Ribault:2005ms, Hikida:2007tq}. It gives a relation of $n$-point functions of the worldsheet theory on $\mathrm{AdS}_3$ with $(2n-2)$-point functions of Liouville theory. The additional operators are degenerate operators of Liouville theory. It would be intriguing to embed the $H_3^+$/Liouville correspondence into the correspondence discussed in this paper.

\paragraph{Checks on the superstring proposal.} We did not perform any quantitative checks on the proposal of the holographic description in the superstring case. It would be important to do so and check that the same matching can be achieved in that case. Moreover, there are more complicated supersymmetric backgrounds than $\mathrm{AdS}_3 \times \mathrm{S}^3 \times \mathbb{T}^4$ that one can investigate.

\paragraph{Higher genus correlators.} We only considered genus 0 correlation functions in the present paper. There are a number of reasons why considering higher genus contributions to the correspondence is interesting. First of all, they are of course only well-defined for the superstring case. Their explicit form on the worldsheet is also unknown and thus studying the correspondence at higher genus could also teach us something about the emergence of the moduli space integral from conformal perturbation theory. Given that the matching at genus 0 was quite intricate and mathematically interesting, the author is all but sure that higher genus will hold more mathematical surprises.

\paragraph{Hidden symmetry.} We want to emphasize that it is surprising that this CFT should have an exact solution. Besides the (super)-Virasoro algebra, it does not have any additional chiral currents, but its (effective) central charge is very large. Usually, conformal symmetry in this regime is not very constraining. It is thus all the more surprising that one is able to resum the conformal perturbation theory answer into the worldsheet expression. To us, this shows that the problem has a hidden symmetry. This hidden symmetry becomes manifest in the string worldsheet descriptions where the $\mathfrak{sl}(2,\mathbb{R})_k$ current algebra leads to powerful constraints. It would be desirable to have a more direct understanding of this symmetry from the CFT perspective. Presumably, the deformation is integrable and one can perhaps use integrability in an effective way to determine the correlators.

\paragraph{Relation to the ELSV formula. } The ELSV formula \cite{ELSV} expresses the Hurwitz number on the sphere with one complicated branch points (specified by a partition) and $m$ more simple branch points (i.e.\ square root branch points) in terms of an integral of certain characteristic classes over moduli space. The Hurwitz number is just counting the number of terms in the sum over covering maps in the symmetric orbifold in eq.~\eqref{eq:deformed correlation function} (and hence one is considering a topological version of the symmetric orbifold). As in our case, the addition of $m$ more simple branch points corresponds to a deformation of the theory by a twist-2 operator. 
The relation described in this paper between string theory on $\mathrm{AdS}_3$ and the deformed symmetric orbifold seems to be a version of the ELSV formula for the physical string. However, the proposals are developed for different regimes: The ELSV formula is formulated for only one insertion of a vertex operator, but at arbitrary genus, whereas we have considered genus 0 and up to three-point functions.

\paragraph{Relation to matrix string theory.} Our proposal is in some ways similar to the proposal of matrix string theory that relates flat space string theory to a deformation of the symmetric orbifold $\mathrm{Sym}^N(\mathbb{R}^{24})$ for the bosonic string \cite{Motl:1997th, Dijkgraaf:1997vv}. See \cite{Arutyunov:1997gt, Arutyunov:1997gi} for a similar computation in this setting than the one we have performed in the present paper. 
In this case, the symmetric product orbifold has to be deformed by an irrelevant operator of dimension $\frac{3}{2}$ given by the twist-2 ground state. The map of parameters is also different. In matrix string theory, the light cone momentum is mapped to the length of the twist divided by $N$ which are then both taken to infinity. In the present case of $\mathrm{AdS}_3$ string theory, we instead keep the twist finite as we send $N \to \infty$. The rough analogue of the light cone momentum is the $\mathrm{SL}(2,\mathbb{R})$ spin $j$, which instead enters explicitly as a momentum in the linear dilaton direction. To relate our description of $\mathrm{AdS}_3$ string theory to the flat space string, one has to take the limit $k \to \infty$, which decouples all the twisted sectors. 
Thus, we see that the two descriptions are in some sense complementary. It would be interesting to understand the connection between the two descriptions of string theory.

\section*{Acknowledgements} I would like to thank Andrea Dei for countless discussions, initial collaboration and comments on the draft of this manuscript. I also thank Matthias Gaberdiel and Emil Martinec for very useful conversations and comments on the draft of this manuscript. I also acknowledge interesting discussions with Shota Komatsu.
This material is based upon work supported by the U.S. Department of Energy, Office of Science, Office of High Energy Physics under Award Number DE-SC0009988
\appendix

\section{Reduction of the three-point function to the two-point function} \label{app:reduction two-point function}

In this appendix, we compute the special worldsheet three-point function with $w_1=w_2=w$, $j_1=j_2=j$, $w_3=2$, $j_3=3-k$, $h_1=h_2=h$ and $h_3=1$.

Let us start with the prefactor $D(j_1,j_2,j_3)$. We find using the functional identities \eqref{eq:Barnes double Gamma function functional identities}
\be 
D(j,j,3-k)=-\frac{\lgamma(2-2j) \lgamma(2k-4)}{2\pi^2 \nu^{2j+2-k} \lgamma\left(\frac{k-1}{k-2}\right)(k-2)^2 \lgamma\left(\frac{2j-1}{k-2} \right) \lgamma(k-2) \lgamma(k-2j-1)}\ . \label{eq:special D prefactor}
\ee
The harder part of the computation is to compute the integral over $(y_1,y_2,y_3)$. After replacing $y_1 \to (-1)^w y_1$ and $y_2 \to -y_2$, the integral reads
\begin{multline}
\int \prod_{i=1}^3 \frac{\mathrm{d}^2 y_i}{\pi} \ \Big| w^{2j+3-k}(y_1 y_2)^{\frac{k w}{2}+j-h-1}y_3 \left((w+1)+y_1+y_2-(w-1)y_1y_2 \right)^{3-2j-k} \\
\times\left(\tfrac{w(w+1)}{2}-\tfrac{w(w-1)}{2} y_1+y_3+y_1y_3 \right)^{k-3} \\
\times\left(-\tfrac{w(w+1)}{2} +\tfrac{w(w-1)}{2}y_2+y_3+y_2 y_3 \right)^{k-3} \Big|^2\ .
\end{multline}
The integrand can be simplified significantly by making the change of variables
\be 
y_3=\frac{w(w+1)+y_1+y_2-(w-1)y_1y_2)}{(1+y_1)(1+y_2)} u+\frac{w((w-1)y_1-w-1)}{2(1+y_1)}\ ,
\ee
which brings the integral into the form
\begin{multline}
\int \prod_{i=1}^2 \frac{\mathrm{d}^2 y_i}{\pi} \frac{\mathrm{d}^2 u}{\pi}\ \Big|\frac{w^{2j-1}}{2}u^{k-3}(1-u)^{k-3}(y_1 y_2)^{\frac{k w}{2}+j-h-1}\left((1+y_1)(1+y_2)\right)^{1-k}\\
\times\left((w+1)+y_1+y_2-(w-1)y_1y_2 \right)^{-2j+k-2} \\
\times\left((1+y_1)(w+1-(w-1)y_2)u-(1+y_2)(w+1-(w-1)y_1)(1-u)\right)\Big|^2\ .
\end{multline}
We can now perform the integral over $u$ explicitly. We have
\begin{align}
\int \frac{\mathrm{d}^2 u}{\pi} \left| u^{k-3}(1-u)^{k-3} (A+B (1-u)) \right|^2=\frac{1}{4}|A+B|^2 \frac{\lgamma(k-2)^2}{\lgamma(2k-4)}\ .
\end{align}
Hence our integral becomes a two-dimensional integral 
\begin{multline} 
\frac{\lgamma(k-2)^2}{4\lgamma(2k-4)} \int \prod_{i=1}^2 \frac{\mathrm{d}^2 y_i}{\pi} \Big|w^{2j}(y_1-y_2)(y_1 y_2)^{\frac{k w}{2}+j-h-1}\left((1+y_1)(1+y_2)\right)^{1-k}\\
\times\left((w+1)+y_1+y_2-(w-1)y_1y_2 \right)^{-2j+k-2} \Big|^2\ . 
\end{multline}
We now perform a further change of variables by setting
\be 
a=\frac{y_1 y_2}{(1+y_1)(1+y_2)}\ , \qquad
b=y_1+y_2\ .
\ee
In these variables the integral simplifies and can be completely evaluated as follows
\begin{multline} 
\frac{\lgamma(k-2)^2}{4\lgamma(2k-4)} \int \frac{\mathrm{d}^2a\, \mathrm{d}^2 b}{\pi^2} \ \left| a^{\frac{kw}{2}+j-h-1}(1-a)^{-2j}b^{2j-1}(1-b)^{k-2j-2} \right|^2\\
=\frac{\lgamma(k-2)^2\lgamma\left(\frac{kw}{2}+j-h\right)\lgamma(k-2j-1)}{4\lgamma(2k-4) \lgamma\left(\frac{kw}{2}-j-h+1\right)\lgamma(k-1)}\ .
\end{multline}
Together with the prefactor \eqref{eq:special D prefactor}, we hence obtain \eqref{eq:special worldsheet 3-point function}.

\section{Poles in the string three-point function} \label{app:string bulk poles}
In this Appendix, we demonstrate that the set of bulk poles of the string three-point functions coincides with the set \eqref{eq:bulk poles} that we motivated from the CFT perspective. The analysis is different depending on the parity of $\sum_i (w_i-1)$.
From the work of \cite{Dei:2021xgh} it is known that the three-point functions are reflection symmetric. Hence the set of poles is automatically reflection symmetric in the three-spins. It thus suffices to look for bulk poles in $j_1+j_2+j_3$.

\paragraph{Even sector.} Let us start with the case $\sum_i (w_i-1) \in 2\mathbb{Z}+1$, in which case the first formula in \eqref{eq:3-point function worldsheet} holds. As mentioned above, we will only look for bulk poles in $j_1+j_2+j_3$.
The Barnes double Gamma function $\G_k(x)$ has the following poles:
\be 
x=m(k-2)+n \ , \qquad m,\, n \in \mathbb{Z}_{\ge 0}\quad \text{or}\quad  m ,\, n \in \mathbb{Z}_{\le -1}\ .
\ee
Consequently, the following poles are manifestly present in the prefactor $D(j_1,j_2,j_3)$ (originating from the term $\G_k(1-j_1-j_2-j_3)$ in the definition \eqref{eq:unflowed three-point function} and already discussed for example in \cite{Maldacena:2001km}):
\be 
j_1+j_2+j_3-1=m(k-2)+n \ , \quad m ,\, n \in \mathbb{Z}_{\ge 1}\quad \text{or}\quad m,\,  n \in \mathbb{Z}_{\le 0}\ .
\ee
This set is not reflection symmetric and we are missing the poles with $m=0$, $n \in \mathbb{Z}_{\ge 1}$ that are present in the CFT answer. In fact, since the worldsheet answer is reflection symmetric they automatically have to be there and come from the integral over the $y_i$'s. We essentially already saw this in the string computation in Section \ref{subsec:string computation} (in the case $m=1$). For $\sum_i j_i =n+1\in \mathbb{Z}_{\ge 2}$, the exponents of $X_{ij}$ sum up to $-n-1 \in \mathbb{Z}_{\le -2}$. This leads to a divergence in the answer, since after the change of variables discussed in \ref{subsec:string computation}, one has to perform an integral
\be 
\int \mathrm{d}^2r\  |r^{-n}|^2\ ,
\ee
near $r=0$, which is divergent even after analytic continuation for $n \in \mathbb{Z}_{\ge 1}$, thus leading to the remaining poles.
\paragraph{Odd parity.} We will next look at the case of even $\sum_i(w_i-1)$. In this case, there is a set poles coming from the factor $\G_k(\frac{k}{2}-j_1-j_2-j_3)$ of the prefactor $D(\frac{k}{2}-j_1,j_2,j_3)$, that gives
\be 
j_1+j_2+j_3-1=m(k-2)+n ,\quad m \le \frac{1}{2}, \ n \le 0 \quad \text{or}\quad m \ge \frac{3}{2},\ n \ge 1\ ,
\ee
where $m \in \mathbb{Z}+\frac{1}{2}$, $n \in \mathbb{Z}$. Comparing to \eqref{eq:bulk poles}, we are missing two sets of poles, namely those with $m=-\frac{1}{2}$ and $n \ge 1$ and those with $m=\frac{1}{2}$ and $n \ge 1$. We already saw in Section~\ref{subsec:string computation} how they arise. For $j_1+j_2+j_3=\frac{k}{2}+n$ for $n \le -1$, the exponent of $X_{123}$ is a negative integer, which leads to the divergence. For $j_1+j_2+j_3=2-\frac{k}{2}+n$, the integral completely localizes on the locus of $y_i=a_i$ as we have seen in the special case of $n=1$ in Section~\ref{subsec:string computation}. The cases for higher $n$ are completely analogous.

\paragraph{Edge cases.} We should mention that this analysis is true in the generic case. There are exceptions in the edge case where
\be 
2\max_i (w_i+1)=\sum_i (w_i+1)\ .
\ee
Let us suppose $w_3=w_1+w_2+1$. Then (up to inconsequential signs)
\be 
X_1=1\ , \quad X_2=1\ , \quad X_3=y_3\ , \quad X_{123}=1+y_1y_3+y_2y_3+\binom{w_1+w_2}{w_1} y_3\ .
\ee
After renaming $y_3 \to y_3^{-1}$ and rescaling the integration variables, we find for the worldsheet three-point function with $w_3=w_1+w_2+1$
\begin{multline}
\left\langle V_{j_1,h_1}^{w_1}(0;0)V_{j_1,h_2}^{w_2}(1;1) V_{j_3,h_3}^{w_3}(\infty;\infty) \right \rangle=\mathcal{N}(j_1) D(\tfrac{k}{2}-j_1,j_2,j_3) \binom{w_1+w_2}{w_1}^{h_3-h_1-h_2}\\
\times \frac{\lgamma\left(j_1+\frac{kw_1}{2}-h_1 \right)\lgamma\left(j_2+\frac{kw_2}{2}-h_2 \right)\lgamma\left(j_3-\frac{kw_3}{2}+h_3 \right)\lgamma\left(h_1+h_2-h_3\right)}{\lgamma\left(j_1+j_2+j_3-\frac{k}{2}\right)}\ .
\end{multline}
While the $\lgamma$-functions in the numerator lead to LSZ-type poles, the structure of the bulk poles is modified by the $\lgamma$-function in the denominator. It cancels those with $m=\frac{1}{2}$ and $n \le 0$ and adds new poles with $m=\frac{1}{2}$ and $n \ge 1$. Thus, the pole structure for the edge case is
\be 
j_1+j_2+j_3-1=m(k-2)+n ,\quad m \le -\frac{1}{2}, \ n \le 0 \quad \text{or}\quad m \ge \frac{1}{2},\ n \ge 1
\ee
with $m \in \mathbb{Z}+\frac{1}{2}$ and $n \in \mathbb{Z}$.

\bibliographystyle{JHEP}
\bibliography{bib}
\end{document}